\documentclass{JHEP}
\usepackage{epsfig}
\usepackage{amssymb}
\usepackage[active]{srcltx}

\newcommand{\be}{\begin{equation}}
\newcommand{\ee}{\end{equation}}
\newcommand{\bea}{\begin{eqnarray}}
\newcommand{\eea}{\end{eqnarray}}
\newcommand{\bean}{\begin{eqnarray*}}
\newcommand{\eean}{\end{eqnarray*}}
\newcommand{\mat}[1]{\left( \matrix{#1} \right)}
\newcommand{\tmat}[1]{{\scriptsize \mat{#1}}}
\newtheorem{theorem}{\sf THEOREM}

\def\IZ{\mathbb{Z}}
\def\IR{\mathbb{R}}
\def\IQ{\mathbb{Q}}
\def\IC{\mathbb{C}}
\def\IP{\mathbb{P}}

\def\ba{\begin{array}}
\def\ea{\end{array}}
\def\beq{\begin{equation}}
\def\eeq{\end{equation}}
\def\braket#1{\left\langle #1 \right\rangle}

\newcommand{\fref}[1]{Figure~\ref{#1}}
\def\eref#1{(\ref{#1})}

\preprint{MIT-CTP-3304\\
 UPR-1012-T\\
{\tt hep-th/0209228}}
\title{Unhiggsing the del Pezzo}
\author{Bo Feng$^1$, Sebasti\'{a}n Franco$^2$, Amihay Hanany$^2$ and Yang-Hui He$^3$,
\\
$^1$Institute for Advanced Study, Princeton, NJ 08540\\ ~\\
$^2$Center for Theoretical Physics,
Massachusetts Institute of Technology,\\ Cambridge, MA 02139, USA\\
~\\ 
$^3$Dept.~of Physics and Astronomy,
        The University of Pennsylvania\\ 
        209, S.~33rd st., Philadelphia, PA 19104.\\ 
\email{fengb@sns.ias.edu,sfranco,hanany@ctp.mit.edu,yanghe@physics.upenn.edu}
}
\abstract{We develop an unhiggsing procedure for finding the D-brane
probe world volume gauge theory for blowups of geometries whose gauge
theory data are known. As specific applications we unhiggs the well-studied
theories for the cone over the third del Pezzo surface. We arrive at
what we call pseudo del Pezzos and these will constitute a first step
toward the understanding of higher, non toric del
Pezzos. Moreover, our methods and results give further support for
toric duality as well as obtaining superpotentials from global
symmetry considerations.}
\begin{document}
\section{Introduction}
D-brane probes to singularities have by now become an important tool
in understanding the compactification of string theory on Calabi-Yau
manifolds. Indeed the resolution of the singularities
\cite{DGM,Aspinwall,DM} to smooth
Calabi-Yau's by the sub-stringy scale dynamics of
the world-volume gauge theories is of great interest to the physicist
and mathematician alike.

With the help of the myriad of combinatorial techniques of toric
geometry, notably the systematic partial
resolution by blowups of Abelian
orbifolds, a particular class of non-compact, singular Calabi-Yau
threefolds have been extensively investigated\footnote{Recently, new phenomenological constructions have been developed 
by wrapping D6-branes on compact, intersecting 3-cycles of Calabi-Yau manifolds \cite{Uranga-rec,blumenhagen}.}
\cite{DGM,Aspinwall,Chris1,DD}. 
These are the so-called toric singularities. Well-studied cases
include Abelian orbifolds and the famous conifold. 
Though the construction of the world-volume gauge
theories for arbitrary singularities which model the Calabi-Yau
remains an open question \cite{Berenstein,finite,BD}, progress has
been made in this subclass.

The method of extracting the world-volume gauge theory of the D-brane
probing transversely to such toric singularities has been formalised
and conveniently algorithmised in \cite{toric}. An interesting
by-product of the so-called Inverse Algorithm is the
phenomenon of Toric Duality where a systematic method has been created
to construct
classes of vastly different gauge theories having the same (toric)
moduli space in the infra-red \cite{toric,phases,seiberg,multiplicity}. A subset 
of the gauge theories that share the same toric moduli space, the {\it toric phases}, have the
interesting property of laying in the conformal window.
Recently,
activities from three different perspectives have hinged on the
conjecture that toric duality is generalised Seiberg's ${\cal N}=1$
duality \cite{seiberg,Beasley:2001zp,CFIKV}.

Prime examples of toric duality and the Inverse Algorithm have been
the cones over del Pezzo surfaces. These surfaces sit as compact
4-cycles (divisors) in the Calabi-Yau and have been a long-time player
in the field of String Theory.
There are in total 10 of such surfaces, namely $\IP^1
\times \IP^1$, $\IP^2$ and $B_k := \IP^2$ blown up at $k=1, \ldots 8$
points. Research of these surfaces in string theory has been diverse
and has ranged over directions from mirror symmetry \cite{HIV} to
mysterious dualities \cite{mystery}. 

The first 5 members of the series, namely the cones over $F_0 := \IP^1
\times \IP^1$, $\IP^2$ (which gives 
the resolution of the famous orbifold
$\IC^3/\IZ_3$) as well as $dP1,2,3$ (the cones of the first 3 del
Pezzo surfaces) are toric and have been
scrutinised in the context of D-brane probes by
\cite{Chris1,toric,phases,seiberg,Beasley:2001zp,CFIKV,multiplicity,soliton,Hanany:2001py},
especially since the advent of the Inverse Algorithm.
The remaining members, $dP4,\ldots,8$, are non-toric and the first
venture into this {\it terra incognita} has been \cite{Hanany:2001py},
wherein the quiver diagrams have been constructed.

Indeed, of late four techniques have been in circulation, towards 
the full
understanding of probing toric singularities: (1) direct field theory
techniques wherein the acquisition of vevs to spacetime fields is
considered \cite{DGM,Aspinwall,Uranga,Chris1,Beasley:2001zp}, (2)
brane configurations such as diamonds \cite{Karch,seiberg} and
$(p,q)$-web techniques \cite{pq,Hanany:2001py,webs,soliton}, (3)
geometric engineering wherein exceptional collections of coherent
sheafs over the divisors provide the gauge theory data and certain geometric
transitions provide large $N$ dualities \cite{CKV,CFIKV,Oh}, 
as well as (4) the Inverse Algorithm
\cite{toric,phases,seiberg,multiplicity}, which is computationally
very convenient and methodical. All these complementary techniques
have thusfar supported each other perfectly, as in particular
exemplified in the detailed study of the above five toric varieties.

However, to have a better understanding of the D-brane probe theory, we
need to proceed beyond toric varieties. In this paper, we 
develop a systematic method, the so called {\bf unhiggsing}
mechanism\footnote{During the preparation of this manuscript, YHH has
	learnt from M.~Wijnholt that the latter's collaboration group
	is also working in this direction and has reached similar
	results.}, to deal with this problem. The basic idea is the
following. Given a singularity $Y$, it is relatively simple to
calculate the quiver diagram (matter content) by the aforementioned
geometric methods. The difficult part is to find the 
corresponding superpotential (for example, by calculating the
mapping among the collections of coherent sheafs). 
Now if we know the quiver and 
superpotential of a singularity $X$ which is the blow down of 
$Y$, we can use the unhiggsing mechanism to get the superpotential
of $Y$ more easily\footnote{There are some subtle points in this
inverse process which we will discuss later.}. 

This above method is of course perfectly adapted to our needs: we have
the quivers of the higher $dPk$'s from \cite{Hanany:2001py}, 
we know that each $dPk$ is the
$\IP^1$-blowup of $dP(k-1)$ and we have the full theories for
$dP0,1,2,3$ from \cite{toric}.
Inspired by this philosophy and armed with this technique, we attempt
at finding the corresponding superpotential of the non-toric
$dP4$ and $dP5$ singularities, with quiver diagrams given in \cite{Hanany:2001py}.
The results turn out to be toric. In other words, 
the moduli space of these gauge theories, unhiggsed from the known
$dP3$ theory, defined by the subsequent superpotentials and
quiver diagrams, are in fact toric varieties. We will see that
these toric moduli spaces are 
not generic, smooth $dP4$ and $dP5$, but degenerate cases with non-isolated
singularities. These singularities we shall call {\bf pseudo del Pezzos}.
These surfaces, which we denote as $PB_k$,  
over which the $PdPk$ are affine cones,
bear close semblance to the del Pezzo surfaces $B_k$
as they are also $\IP^2$ blown up at points . 

Although we do not reach our
initial aim, the method itself is very useful and can be applied to
hosts of examples in order to construct new classes of D-brane gauge
theories. We will discuss more about this issue in the conclusion.

Furthermore, continuing along the path of
\cite{multiplicity,soliton}, we shall use elegant symmetries
inherited from the very geometry (and indeed from the closed string
sector), to arrive at the superpotentials for these theories \footnote{The issue
of multiplicity symmetry, raised in \cite{multiplicity}, has also been considered
in \cite{Muto}.}. Once
again, we shall find that such symmetry considerations are powerful
enough to uniquely determine the superpotential, the calculation of
which is often a daunting task, either for the Inverse Algorithm, or
for the composition of Ext's in the derived category of coherent
sheafs.

The organization of the paper is as follows. In Section 2, we refresh
the readers' memory on the four toric phases of the $dP3$ theories,
known to the literature. Then, in Section 3, we present the other
ingredient and explain the (un)higgsing mechanism in relation to
geometric blow (down) ups. Thus prepared, we unhiggs the $dP3$
theories to obtain the $dP4$ gauge theory in Section 4, and check the
consistency by higgsing back to $dP3$ in Section 5. As a hind-sight,
in Section 6, we shall see that we have in fact obtained the $PdP4$
theory and discuss some of the geometric properties
thereof. Continuing in this vein, we obtain the $PdP5$ theory in
Section 7. As an additional confirmation to the unhiggsing method, we
also use global symmetry arguments to check our superpotentials
in Section 8. Finally, we conclude in Section 9.
\section*{Nomenclature}
Unless otherwise stated, we shall throughout the paper adhere to the
notation that $dPk$ means the affine cone over the $k$-th del Pezzo
surface $B_k$, i.e., $\IP^2$ blown up at $k$ generic points. When
these blowup points are {\it not} generic, i.e., 3 or more may be
colinear, or 6 or more may lie on a single conic, we shall call the
resulting surface the non-generic (or Pseudo) del Pezzo, denoted as
$PB_k$; some of these may actually be toric as we shall see. The
affine cones over these surfaces we shall call $PdPk$.

Often we shall
append a Roman numeral subscript as in $dPk_I$; this means the $I$-th
(toric-dual) phase of the theory for $dPk$. And so likewise for
$PdPk$.

In the quiver theory,
the arrow $X_{ij}$ corresponds to the bifundamental field from node
$i$ to $j$. 
%

\section{The Four Phases of $dP3$}
\EPSFIGURE[h]{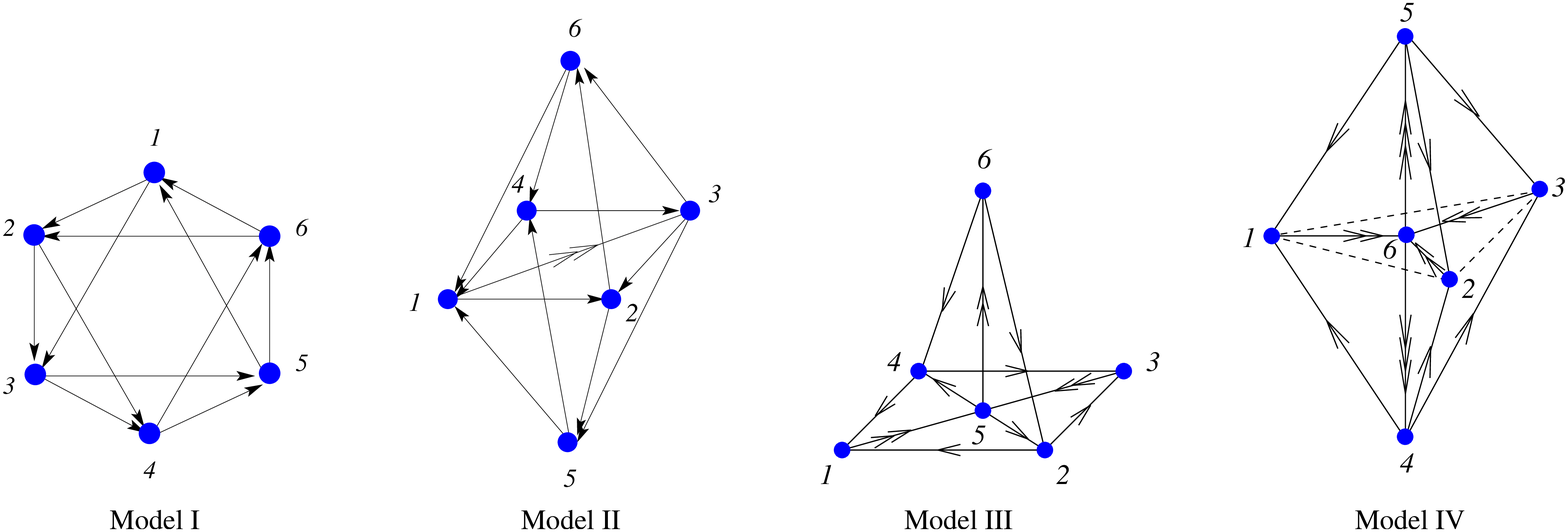,width=19cm}
{Quiver diagrams of the four phases of $dP3$.
\label{quivers_dP3}
}

The starting point for the unhiggsing process that we will use to generate the
theories associated to higher del Pezzos is $dP3$. There are four toric phases 
corresponding to $dP3$ \cite{seiberg,multiplicity,soliton,Beasley:2001zp}. 
To refresh the reader's memory, let us clarify what we mean by a {\bf toric
phase}, as inspired by the Toric Duality discussions in \cite{toric,seiberg}: 
we call any gauge theory where the quiver has the rank of all
nodes equal to $N$ (for simplicity, most times we set $N=1$) 
as well as only monomial F-terms, i.e., suitable for
the Forward Algorithm of \cite{toric,DD}.
Indeed this is not a necessary condition for the moduli space to be toric. We can have
phases without all the ranks of the nodes equal, and still obtaining a toric moduli space when
calculating it in terms of gauge invariant operators.

Now, let us recall the $dP3$
quivers in \fref{quivers_dP3}, where we 
have used the versions presented in \cite{multiplicity}, which make
global symmetries explicit.

The superpotentials for these theories are
\begin{eqnarray}
W_I & = & X_{12} X_{23} X_{34} X_{45} X_{56} X_{61}
-[X_{23} X_{35} X_{56} X_{62}+
X_{13} X_{34} X_{46} X_{61}
+X_{12} X_{24} X_{45} X_{51}] \\ \nonumber 
&&+[X_{13} X_{35} X_{51}+X_{24}
  X_{46} X_{62}]
\label{model1}\\
W_{II} & = & [X_{12} X_{26} X_{61}-X_{12} X_{25} X_{51}+
X_{36} X_{64}X_{43}-X_{35} X_{54}X_{43}] \\ \nonumber 
& & +[-X_{61} X_{13} X_{36}+X_{51} Y_{13} X_{35}]
+[-X_{26} X_{64} X_{41} Y_{13} X_{32}+X_{25} X_{54} X_{41} X_{13}
  X_{32}] 
\label{model2} \\
W_{III} & = & [X_{41} X_{15} X_{54}- X_{54} X_{43} X_{35}
  +Y_{35} X_{52} X_{23}-X_{52} X_{21} Y_{15}] \\ \nonumber 
& & +[-X_{41} Y_{15} X_{56} X_{64}+X_{64} X_{43} Y_{35} Y_{56}
-X_{23} X_{35} X_{56} X_{62}+X_{62} X_{21} X_{15} Y_{56}] 
\label{model3}\\
W_{IV} & = & [X_{41} X_{16} X_{64} + X_{43} X_{36} Y_{64}+X_{42} X_{26} Z_{64}]
- [X_{41} Y_{16} Y_{64} + X_{43} Y_{36} Z_{64}+X_{42} Y_{26} X_{64}] \\ \nonumber 
& + &  [X_{51} Y_{16} X_{65} + X_{53} Y_{36} Y_{65}+X_{52} Y_{26} Z_{65}]
- [X_{51} X_{16} Y_{65} + X_{53} X_{36} Z_{65} +X_{52} X_{26} X_{65} ] 
\label{model4}.
\end{eqnarray}

From this data, we shall use the technique of ``unhiggsing''
to attempt to arrive the theories for the higher del Pezzos.

%
\section{Blowing Up and Down versus Unhiggsing and Higgsing}

Now we need our second ingredient and discuss the geometric origin of
the (un)higgsing method. The philosophy is straight-forward and
standard to the literature:

{\it
the blow-up of a point, replacing it by a compact 2-cycle, is translated to
an unhiggsing of the field theory on the D-brane. Conversely, blowing
down a 2-cycle corresponds to the higgsing of turning on a VEV for a bifundamental
field that breaks two $U(1)$ factors down to a single one.
}

In terms of fractional branes, the higgsing process corresponds to the
combination of the
fractional branes of the higgsed gauge groups into bound states as
discussed in \cite{webs}.

Let us now discuss the connection between the higgsing and the partial
resolution methods
\cite{Beasley:2001zp,toric,Chris1,DD}.
When Fayet-Iliopoulos (FI) terms acquire generic values the singularity is completely
resolved. On the other hand, when the FI terms
lie on some non-generic cones, we obtain a partial resolution
corresponding to a non-trivial (singular)
geometry. This technique was exploited in
\cite{DGM,Aspinwall,Uranga,Chris1,toric} 
to obtain theories for various toric
varieties starting from abelian orbifolds. To illustrate, let us
consider the resolution of the  
$\IC^3/(\IZ_2 \times \IZ_2)$ down to the Suspended Pinched Point (SPP). The quiver for 
$\IC^3/(\IZ_2 \times \IZ_2)$ is given in \fref{f:z2z2toSPP}(a) (which we quote from
\cite{Uranga,toric}), while its superpotential is
\begin{eqnarray}
W=X_{13} Y_{34} Z_{41}-X_{13} Z_{32} Y_{21}+X_{31} Y_{12}
 Z_{23}-X_{31} Z_{14} Y_{43} \\ \nonumber 
 +X_{24} Y_{43} Z_{32}-X_{24} Z_{41} Y_{12}+X_{42} Y_{21} Z_{14}-X_{42} Z_{23} Y_{34}.
\end{eqnarray}

\EPSFIGURE[ht]{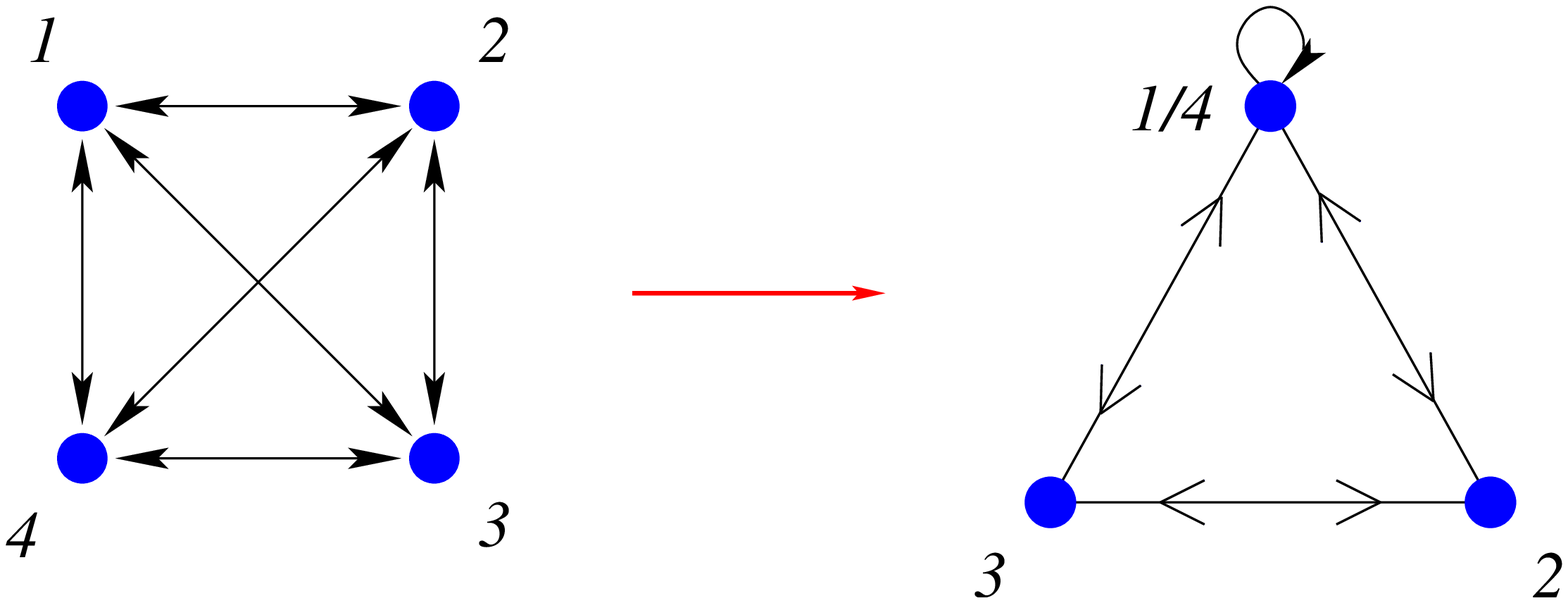,width=10cm}
{(a) The quiver for the parent orbifold $\IC^3/(\IZ_2 \times \IZ_2)$;
  (b) The quiver for the SPP, a partial resolution from higgsing
the parent.
\label{f:z2z2toSPP}
}

The SPP is obtained by constraining the four FI terms to be \cite{Uranga,toric}
\beq
\label{FI_relations}
\begin{array}{cccc}
\zeta_2=0 \ \ & \ \ \zeta_3=0 & \ \ \zeta_1+\zeta_4=0 \ \ &  \ \ \zeta_1 \neq 0.
\end{array}
\eeq

This corresponds to higgsing $U(1)_{(1)} \times U(1)_{(4)}$ to a
single $U(1)$. We can do it
by giving a non-zero VEV to $Z_{14}$ (the alternative of giving a VEV to 
$Z_{41}$ is equivalent by symmetry). Let us set $\langle Z_{14}
\rangle = 1$. 
During the higgsing process, mass terms are generated for $X_{31}$,
$Y_{43}$, $X_{42}$ and $Y_{21}$, so they have to be integrated
out. Calling nodes $1(4) \rightarrow 1$, we get
\beq
W=X_{21} Y_{12} Z_{23} Z_{32}-Z_{32} Z_{23} Y_{31} X_{13}+X_{13} Y_{31} Z_{11}-
X_{21} Z_{11} Y_{12}
\eeq
and the quiver in \fref{f:z2z2toSPP}(b), which is exactly that for the SPP.

More explicitly, let us consider the D-terms of nodes $1$ and $4$.
If we give only one field $Z_{14}$ a nonzero VEV, to satisfy D-terms for
these two nodes, both $\zeta_1$ and $\zeta_4$ can not be zero, but 
$\zeta_1+\zeta_4=0$ because of the opposite sign of field $Z_{14}$ in these
two D-terms. This establishes the relationship between FI-parameters
and  fields which acquire nonzero VEV.

Therefore, we have shown in a simple example how the linear relations among
FI parameters associated 
to a blow-down such as \eref{FI_relations} straightforwardly determine
a higgsing in the gauge theory.
The methodology is of course easily generalised and the reverse of the
procedure, viz., the unhiggsing is much in the same spirit and will be
detailed in the next section. We remark that such relation between
(un)higgsing and blowing (up) down is very conveniently visualised in
the $(p,q)$-web picture \cite{pq,webs}.
%
\section{Unhiggsing From $dP3$ to $dP4$}
\label{sec:unhiggs}
Thus our ingredients are complete. With the full theories for $dP3$,
the quivers for the higher (non-toric) del Pezzo's
given in \cite{Hanany:2001py}, as well as the
preparatory \'{e}tude on the SPP in the previous section, let us
proceed.

The quiver diagram of $dP4$ given in \cite{Hanany:2001py} is redrawn
here as Model I in \fref{f:dP4I}. The other models we shall obtain later.
For this phase of $dP4$, we have a total of 15 fields. When we higgs
down to $dP3$ in the manner of Section 3 therefore, we can
 reach at most three of total four phases, 
viz.~$dP3_I$ with 12 fields as well as
$dP3_{II}$ and $dP3_{III}$ with 14 fields. 
For $dP3_{IV}$ there are 18 fields so it obviously can not be
higgsed from this phase of $dP4$. 
Let us analyze this process in more detail.
\EPSFIGURE[ht]{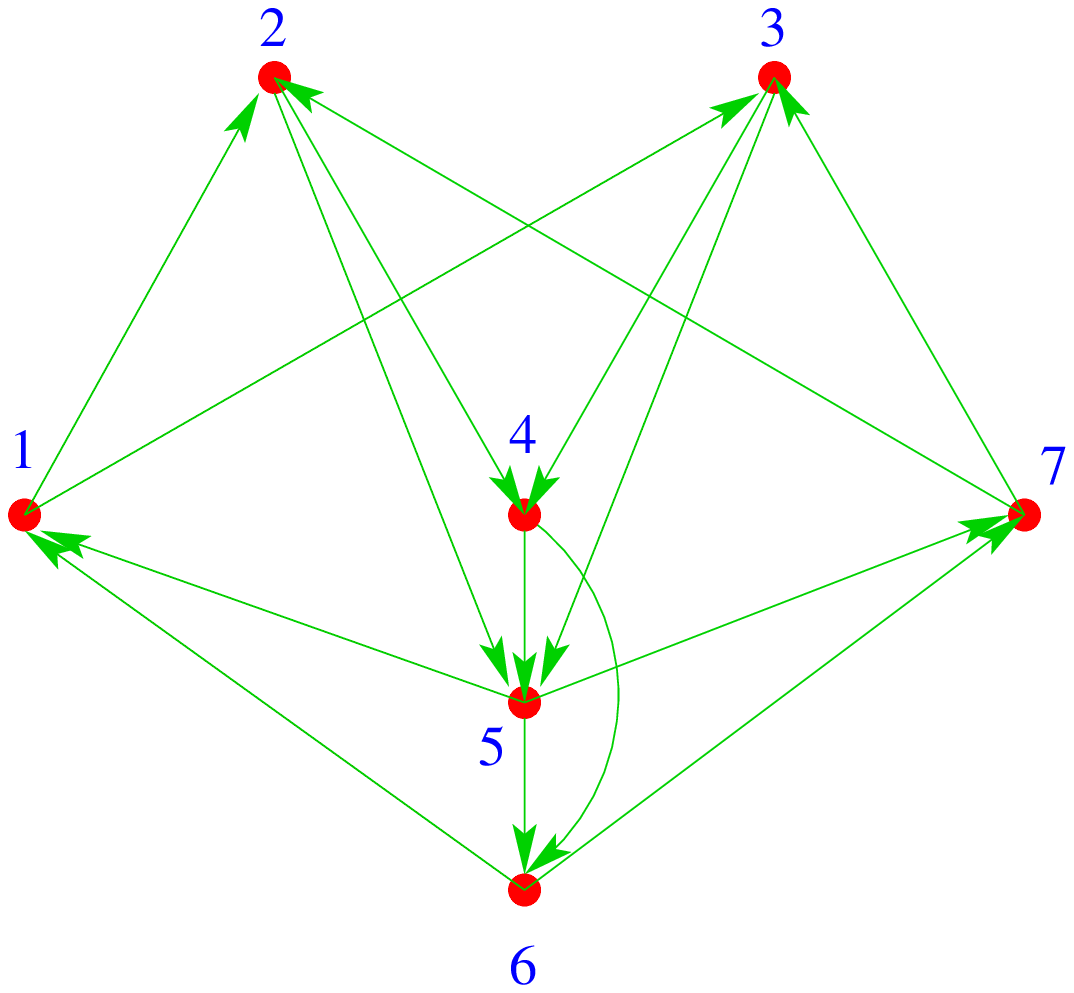,width=7.5cm}
{The quiver diagram for $dP4_I$, redrawn from the results in
\cite{Hanany:2001py}. In this paper this model is referred to as Model
I, a $U(1)^7$ theory with 15 bifundamentals.
\label{f:dP4I}
} 
\subsection{Higgsing $dP4_I$}
First, notice that there is an explicit symmetry of the quiver for
$dP4_I$, namely the reflection about the 456-axis.
This means that node-pair $(2,3)$ as well as the pair $(1,7)$ are
equivalent. Now let us see whether $dP4_I$ can be higgsed down to
$dP3_I$; the latter seems a natural choice because it, as with
$dP4_I$, is the only model without multiple arrows between any two
nodes.
However, we can not find the reflection symmetry exhibited in $dP4_I$,
i.e., we can not find such equivalence between pairs of nodes in $dP3_I$.
This tells us that when we higgs down, such symmetry could be broken
and in fact will be so.

Second, notice that for node $5$ in  $dP4$, we have three arrows
coming in and
three going out while there are only two incoming and two outgoing
arrows for any node in $dP3_I$. This means that we must integrate out 
one incoming 
and one outgoing arrow at node $5$ when we higgs down;
these two fields must acquire mass when we higgs. In other words, there
must be a cubic term in the superpotential that involve these two fields
and another field to which we will give nonzero vacuum expectation
value (VEV).

To set some notations, we shall label fields
in $dP4$ by $\phi$ and those in $dP3$ by $X$. Moreover, in the quiver
diagrams, the daughter of the higgsing will have its nodes indexed by
numbers with ``[~]'' around them and node $a/b [c]$ would thus mean
node $c$ in the daughter, obtained from higgsing nodes $a$ and $b$ in
the parent.

Combining the above two observations, we see that the one field which is
integrated out must be $\phi_{25}$.
Indeed we can make this choice due to the symmetry between 
nodes $2$ and $3$.
Now to get the cubic term which includes $\phi_{25}$, we have
only two choices: $ \phi_{25} \phi_{57} \phi_{72}$ and
$ \phi_{25} \phi_{51} \phi_{12}$ as they are the only closed loops in
the quiver diagram (i.e. gauge invariant operators) involving node
5. Again, since node $7$ is symmetric to  
node $1$, these two choices are equivalent to each other. So without
loss of generality we take $\phi_{25} \phi_{51} \phi_{12}$.

\EPSFIGURE[ht]{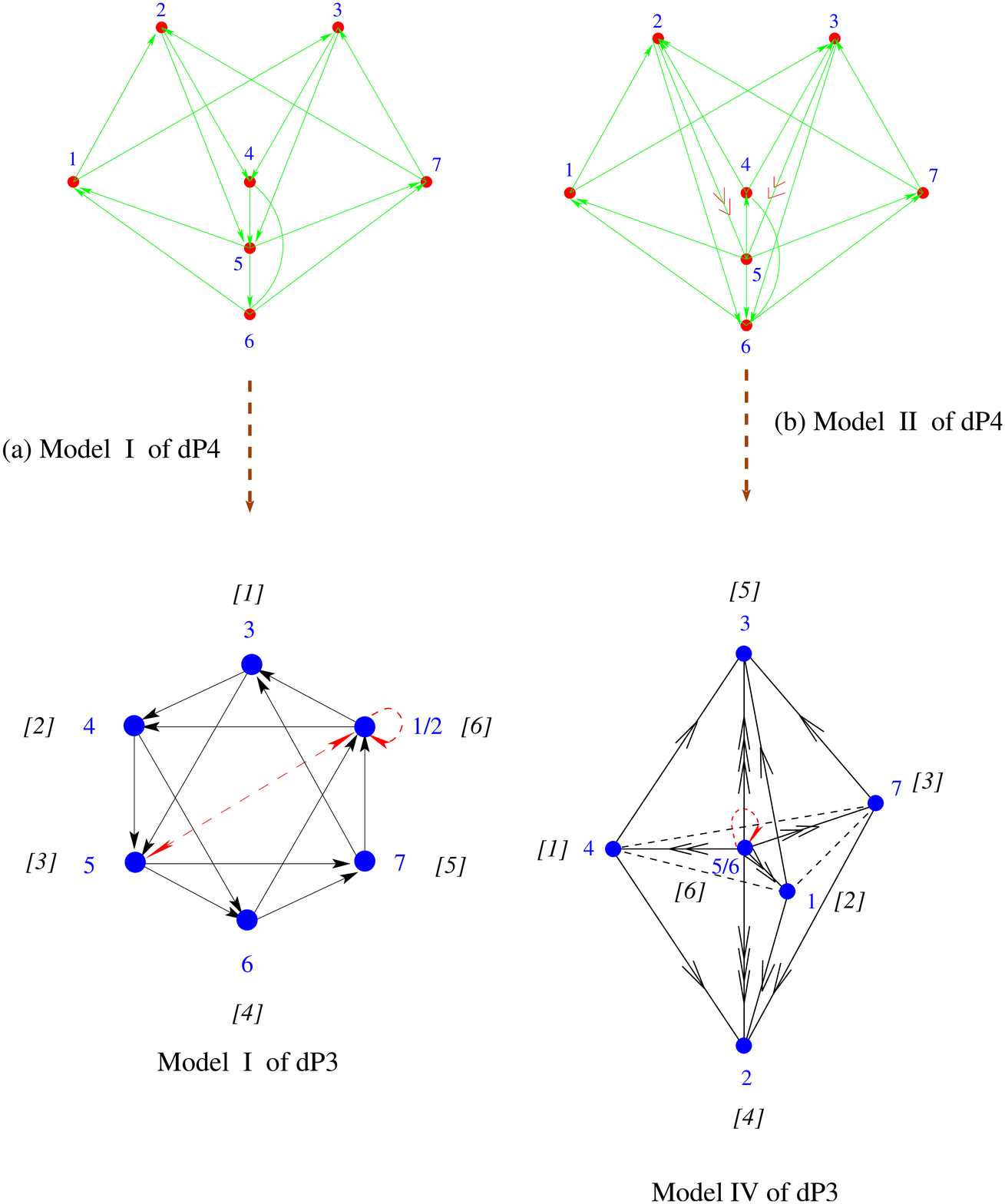,width=6in}
{The higgsing of $dP4$ down to $dP3$.
(a) The quiver diagram after we condense $\phi_{12}$ from $dP4_I$ to
$dP3_I$; (b) similarly we obtain $dP3_{IV}$ from $dP4_I$ by turning on a VEV
for $\phi_{56}$.\label{f:dp4todp3}
}

We should also condense field $\phi_{12}$ from 
$\phi_{25} \phi_{51} \phi_{12}$. The condensation process is shown in
part (a) of \fref{f:dp4todp3} wherein we give the corresponding nodes
of $dP3$ in brackets for comparison. 
We have drawn dashed lines for the field that gets a non-zero VEV and
those that become massive and are integrated out. 

So far we have used the quiver alone, the next step is to start from 
the superpotential of $dP3_I$ to attempt to reach that for $dP4_I$
which is thusfar unknown in the literature. From the superpotential
for $dP3_I$,
\bean
W_{dP3_I} 
& = & X_{12}X_{23}X_{34}X_{45}X_{56}X_{61}-[X_{23}X_{35}X_{56}X_{62}+
X_{13}X_{34}X_{46}X_{61} \nonumber \\
  & & +X_{12}X_{24}X_{45}X_{51}]+[X_{13}X_{35}X_{51}+X_{24}X_{46}X_{62}]
\eean
we replace by the corresponding fields in $dP4$ (in the way that is suggested
by \fref{f:dp4todp3}.a) to get
\bean
W_{dP4_I}
& = & \phi_{34} \phi_{45} \phi_{56} \phi_{67} \phi_{72} \phi^{-1}_{12} \phi_{13} - [\phi_{45} \phi_{57} \phi_{72} \phi_{24}+ \phi_{35} \phi_{56} \phi_{61} \phi_{13} \\ 
\nonumber
 & & + \phi_{73} \phi_{34} \phi_{46} \phi_{67} ] +  [\phi_{73} \phi_{35} \phi_{57}+ \phi_{24} \phi_{46} \phi_{61} \phi_{12}]
\eean

In $W_{dP4_I}$
to close the loops we needed to replace $X_{46}$ by $\phi_{61}
\phi_{12}$. The crucial step is that the term 
$X_{12}X_{23}X_{34}X_{45}X_{56}X_{61}$ must be replaced by
$\phi_{34} \phi_{45} \phi_{56} \phi_{67} \phi_{72} \phi^{-1}_{12}
\phi_{13}$ where we have put in $\phi^{-1}_{12}$ to show
that this term is the result of 
integrating out massive fields. In other words, this term should come
from the replacement of $\phi_{25}$ or $ \phi_{51}$ by their
equations of motion.

If it came from the replacement of $\phi_{25}$, we should have the
term $\phi_{25} \phi_{56} \phi_{67} \phi_{72}$ which upon substitution
of $\phi_{25}$ from
$\phi_{12}\phi_{25}=\phi_{13}\phi_{34} \phi_{45}$ gives the required
$\phi_{34} \phi_{45} \phi_{56} \phi_{67} \phi_{72} \phi^{-1}_{12}
\phi_{13}$. Thus we get the final superpotential as
\bea
\label{superdP4I}
W_{dP4_I} 
& = & \phi_{24} \phi_{46} \phi_{61} \phi_{12}+ \phi_{73} \phi_{35} 
  \phi_{57}
- \phi_{73} \phi_{34} \phi_{46} \phi_{67}-\phi_{45} \phi_{57}
\phi_{72} 
\phi_{24} - \phi_{35} \phi_{56} \phi_{61} \phi_{13} \nonumber \\
& &  +\phi_{25} \phi_{56} \phi_{67} \phi_{72}- 
  \phi_{51}(\phi_{12}\phi_{25}-\phi_{13}\phi_{34} \phi_{45}).
\eea

If on the other hand we were to do the
replacement of $\phi_{51}$, we should have the term
$\phi_{51} \phi_{13}\phi_{34} \phi_{45}$, with the EOM
$\phi_{51}\phi_{12}= \phi_{56} \phi_{67} \phi_{72}$.
From this we would have the superpotential
\bean
W_{dP4_I} 
& = & \phi_{24} \phi_{46} \phi_{61} \phi_{12}+ \phi_{73} \phi_{35} 
  \phi_{57}
- \phi_{73} \phi_{34} \phi_{46} \phi_{67}-\phi_{45} \phi_{57} \phi_{72}
  \phi_{24} - \phi_{35} \phi_{56} \phi_{61} \phi_{13} \nonumber \\
& & +\phi_{51} \phi_{13}\phi_{34} \phi_{45}-\phi_{25}(\phi_{51}\phi_{12}-
 \phi_{56} \phi_{67} \phi_{72}).
\eean
This is the same as \eref{superdP4I}.
 We have therefore obtained the superpotential for $dP4_I$.

\subsection{The Various Phases of $dP4$}

Having obtained one phase of the $dP4$ theory, it is natural to seek
other phases related thereto by Seiberg duality. 
In this section, we shall look for the duality transformations 
which stay within the toric phase.
We shall also find the closure of this set of
dual theories.

For $dP4_I$, we can rewrite \eref{superdP4I} as
\bea
\label{dp4_1}
W_{dP4_I} & = & -[\phi_{51}\phi_{12}\phi_{25}- \phi_{57}\phi_{73} \phi_{35} ]+
  [\phi_{12}\phi_{24} \phi_{46} \phi_{61}-\phi_{73} \phi_{34} 
  \phi_{46} \phi_{67}] \\
& + & [\phi_{34} \phi_{45}\phi_{51}\phi_{13}-\phi_{24}\phi_{45} 
 \phi_{57} \phi_{72}]+[\phi_{25} \phi_{56} \phi_{67} \phi_{72}-
 \phi_{35} \phi_{56} \phi_{61} \phi_{13}]\nonumber
\eea
where to show explicitly the symmetry between the pair $(1,2)$ and $(7,3)$
we have redefined the fields and grouped them properly. We remind the
reader that this can be
higgsed down to model I of $dP3$. 

Now let us discuss the symmetries of this model in the spirit of
\cite{multiplicity}. 
First, from the quiver diagram in \fref{f:dP4I}
we see following symmetry: (1) nodes $1\leftrightarrow 7$; (2)
nodes $2\leftrightarrow 3$; (3) nodes $4\leftrightarrow 6$, 
$1\leftrightarrow 3$, $2\leftrightarrow 7$ as well as reversing the 
directions of all arrows.

However, the superpotential we found in \eref{dp4_1} does not preserve
all these symmetries. It is easy to see that only the following
symmetries are preserved: (1) simultaneous exchange of nodes 
$1\leftrightarrow 7$ and $2\leftrightarrow 3$ (we have
shown this symmetry explicitly by the brackets in \eref{dp4_1}); 
(2) exchange of nodes $4\leftrightarrow 6$, 
$1\leftrightarrow 3$, $2\leftrightarrow 7$ and reversal of the 
directions of all arrows. 

These observations of symmetries are very 
important and will reduce much computation in tracing through the tree
of generalised Seiberg dualities. 
For example, we see immediately that dualising on node $4$ will give
the same theory as on $6$. Similarly, dualising on any of $1,2,3,7$ will
also produce the same theory.

Now starting from $dP4_I$ we can dualise either node
$4,6$ to give us a new model which we shall call
$dP4_{II}$. The superpotential is after integrating out, given by
\bea
\label{dp4_2}
W_{dP4_{II}} 
& = & -[\phi_{51}\phi_{12}\phi_{25}- \phi_{57}\phi_{73} \phi_{35} ]+
  [\phi_{12}\widetilde{\phi}_{26} \phi_{61}-\phi_{73} 
   \widetilde{\phi}_{36}\phi_{67}] \\
& + & [\widetilde{\phi}_{35}\phi_{51}\phi_{13}-\widetilde{\phi}_{25}
 \phi_{57} \phi_{72}]+[\phi_{25} \phi_{56} \phi_{67} \phi_{72}-
 \phi_{35} \phi_{56} \phi_{61} \phi_{13}]\nonumber \\
& + & [\widetilde{\phi}_{25} \phi_{54} \phi_{42} 
  -\widetilde{\phi}_{35} \phi_{54} \phi_{43}]-[\widetilde{\phi}_{26} 
  \phi_{64}\phi_{42}-\widetilde{\phi}_{36} \phi_{64}\phi_{43}]\nonumber,
\eea
where the $\widetilde{\phi}$ are dual meson fields and the last row
comes from the added meson interaction of the form $Mq\tilde{q}$.

Now let us discuss the symmetries of $dP4_{II}$, which from the quiver we
see as (1) $2\leftrightarrow 3$ and (2) the permutations of
nodes $(1,4,7)$. Again, the superpotential preserves only the
symmetry of exchanging $1\leftrightarrow 7$ and $2\leftrightarrow 3$
at the same time. This is also explicitly shown in \eref{dp4_2} by
grouping the appropriate terms in
brackets. The symmetry indicates that dualising nodes $1,7$ will give
the same
theory. It is also worth to mention that although $\phi_{35}$
and $\widetilde{\phi}_{35}$ are doubly degenerate in the quiver
diagram, the superpotential breaks this degeneracy explicitly. 
The same conclusion holds for fields $\phi_{25}$ and
$\widetilde{\phi}_{25}$.

Finally, we have nodes $1,2,3,7$ to choose from in dualising $dP4_I$.
Let us without loss of generality choose to dualise node $1$;
we reach yet another model, which we call $dP4_{III}$. Comparing
with the quiver of $dP4_I$, we notice that they are almost the
same except
one thing: there is a bi-directional arrow between nodes $3,5$.
This difference is very important and non-trivial. For all del Pezzo
surfaces
we have encountered before, they are always chiral. This property of
the del Pezzo surfaces was also pointed out in \cite{Hanany:2001py}.
In fact, the rules given in \cite{seiberg,soliton} about Seiberg
duality can not be directly applied to such cases. 
It is certainly worth to investigate this and generalise the Seiberg
duality rules. 
In any event we seem to have a puzzle here as the
the del Pezzo surfaces admit only uni-directional arrows
\cite{Hanany:2001py}.
We shall address this puzzle in Section \ref{sec:toric}. For
now let us present the superpotential:
\bea 
\label{dp4_3} W_{dP4_{III}} & = & [\phi_{62} \phi_{24} \phi_{46}-\phi_{62}
  \phi_{21} \phi_{16}]+[\phi_{34} \phi_{45} \phi_{53} -\phi_{31} 
 \phi_{15} \phi_{53}] \\
& + & \phi_{57} \phi_{73} \phi_{35} -\phi_{35} \phi_{56} \phi_{63} +
  \phi_{63} \phi_{31} \phi_{16} \nonumber \\
& - & \phi_{73} \phi_{34} \phi_{46} \phi_{67} -\phi_{24}\phi_{45} \phi_{57}
  \phi_{72} +  \phi_{21} \phi_{15}  \phi_{56}  \phi_{67}\phi_{72}
 \nonumber.
\eea
The quiver of this model has only an explicit $\IZ_2$ symmetry 
$1\leftrightarrow 4$.

These three models are the only toric phases of
$dP4$ under Seiberg duality and we summarise them in \fref{f:dP4all}.
\EPSFIGURE[ht]{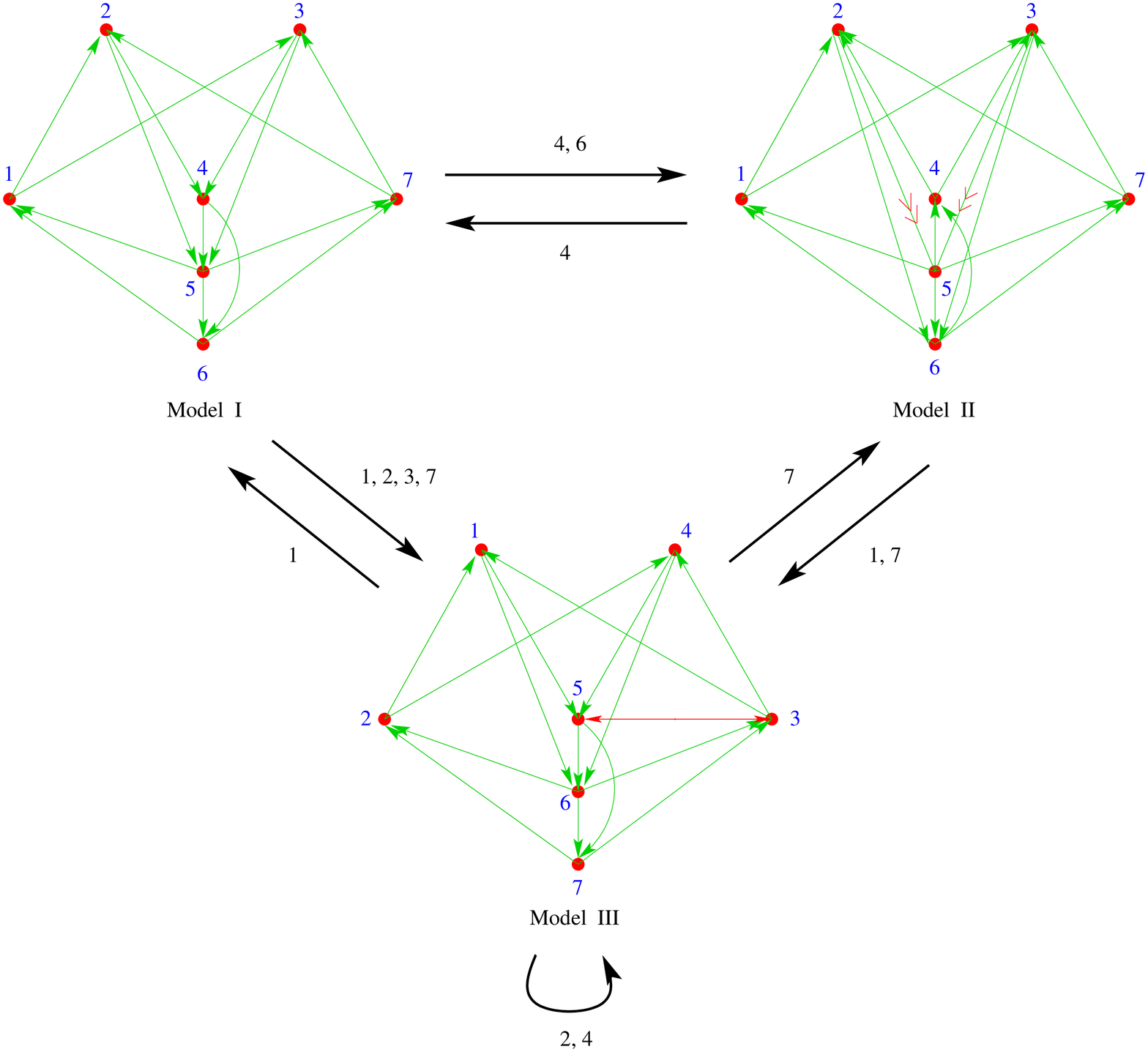,width=15cm}
{The quivers for the three toric Seiberg dual phases of $dP4$.
The nodes upon which one dualises to transform between
them are shown next to the arrows.
\label{f:dP4all}
}
%

\section{Higgsing from $dP4$ to $dP3$}
In the previous sections we have studied how to obtain one of the
phases of $dP4$ by unhiggsing $dP3_I$, and then calculated
all the three toric phases of $dP4$ that are closed under Seiberg
dualities.
Now we will show how it is possible to get all
the four toric $dP3$ phases by higgsing the $dP4$ models.  One can 
conversely adopt the unhiggsing perspective, and think about the
result we here present as possible ways of 
going from $dP3$ to $dP4$ by suitable unhiggsings. Once again one
could take the $(p,q)$ perspective \cite{webs} to visualise more
easily. 
\subsection{Phase I of $dP3$}
The discussion in Section \ref{sec:unhiggs} showed how one
obtains $dP3_I$ from $dP4_I$ and vice versa. 
Here we show how to accomplish the same
using $dP4_{III}$ as our starting point. 

Let us turn on a non-zero 
VEV for $\phi_{31}$ in $dP4_{III}$. This expectation value for a
charged 
bifundamental field breaks $U(1)_{(1)}\times U(1)_{(3)}$
down to the $U(1)_{[5]}$
subgroup, thus leading to a theory with $U(1)^6$ gauge group. 
The subsequent quiver diagram is shown in \fref{dp4todp3_1}.
Looking at the superpotential \eref{dp4_3} we see that the cubic terms 
containing  $\phi_{31}$ give rise to masses for $\tilde\phi_{53}$, 
$\phi_{15}$, $\tilde\phi_{26}$ and $\phi_{16}$. When looking at the IR 
limit of the gauge theory, these massive fields have to be integrated out 
using their equations of motion. The result, is a $U(1)^6$ gauge theory
with 12 fields and superpotential given by
\bean
W=\tilde\phi_{62} \phi_{24} \phi_{46}+\phi_{57} \phi_{73}
\phi_{35}+\phi_{21} \phi_{34} \phi_{45} \phi_{56} \phi_{67} \phi_{72}
\\ \nonumber 
-\phi_{35} \phi_{56} \tilde\phi_{62} \phi_{21}-\phi_{73} \phi_{34}
\phi_{36} \phi_{67}-\phi_{24} \phi_{45} \phi_{57} \phi_{72},
\eean
which, after the following renaming of the gauge groups $(1/3,2,4,5,6,7) 
\rightarrow (5,4,6,1,2,3)$ and calling the fields $X$ and setting 
$\braket{\phi_{31}}=1$ becomes
\bean
W=X_{24} X_{46} X_{62}+X_{13} X_{35} X_{51}+X_{45} X_{56} X_{61}
X_{12} X_{23} X_{34} \\ \nonumber 
-X_{51} X_{12} X_{24} X_{45}-X_{35} X_{56} X_{62} X_{23}-X_{46} X_{61}
X_{13} X_{34}.
\eean
We recognise this exactly as the superpotential, and part (b) of
\fref{dp4todp3_1}, the quiver, for phase I
of $dP3 $, as is required.
\EPSFIGURE[h]{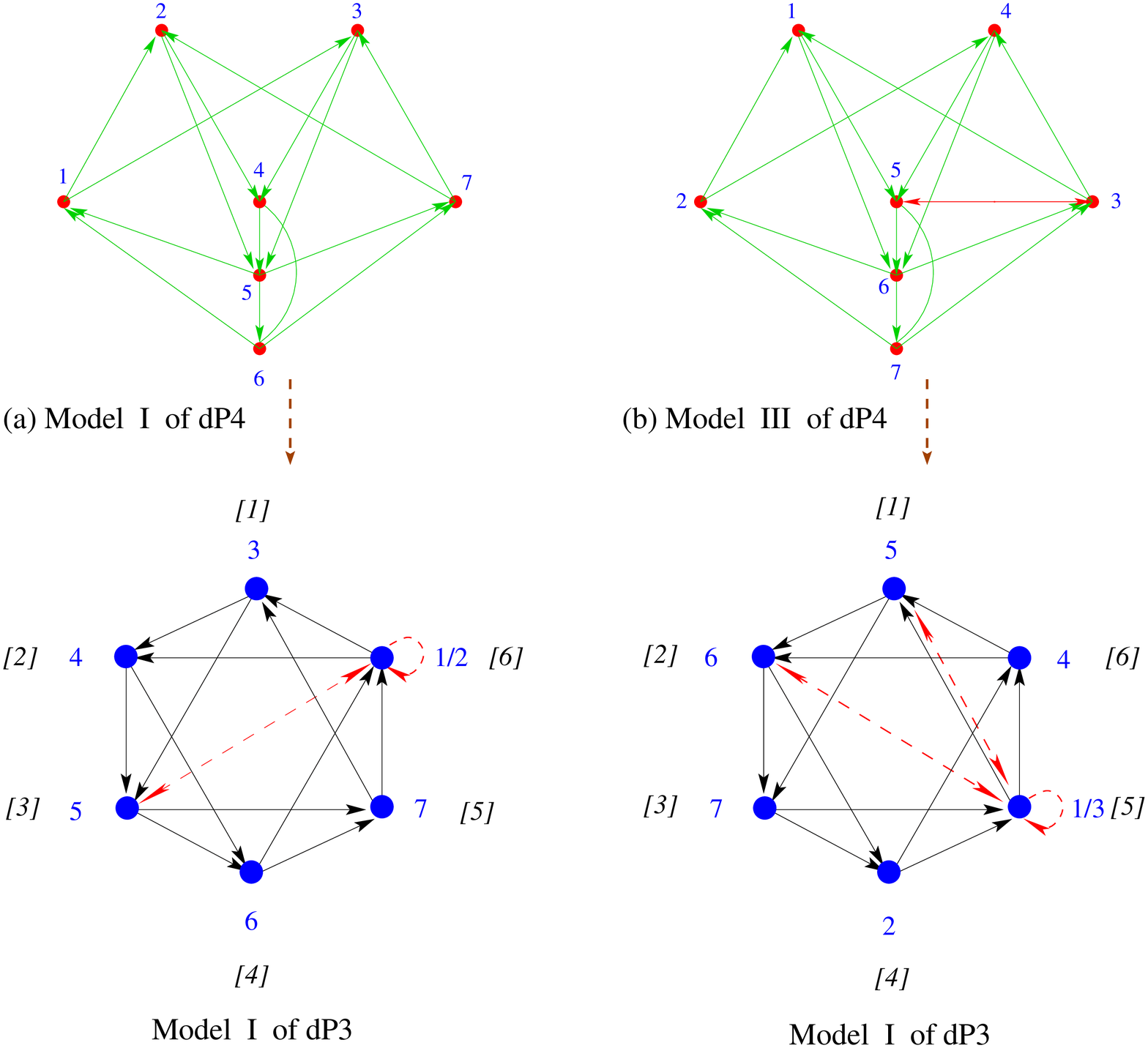,width=15cm}
{(a) Higgsing the field $\phi_{12}$ of $dP4_I$ to obtain $dP3_I$. 
(b) Higgsing the field $\phi_{31}$ of $dP4_{III}$ to also reach
$dP3_I$. 
We have used 
the dashed lines to indicate the fields to be integrated out and
nodes in 
bracket to indicate the corresponding nodes in model I of $dP3$. 
\label{dp4todp3_1}
}
\subsection{Phase II of $dP3$}
Referring to \fref{dp4todp3_2},
let us start from model II of $dP4$ and give a VEV to $\phi_{12}$. 
In this case, only the $U(1)_{[4]}$ in $U(1)_{(1)}\times U(1)_{(2)}$ 
survives. Mass terms are generated 
for $\phi_{25}$, $\phi_{51}$, $\tilde\phi_{26}$ and $\phi_{61}$. 
After integrating them out, we have a $U(1)^6$ theory with 14 fields 
and the following superpotential
\bean
W=\tilde\phi_{36} \phi_{64} \phi_{43}+\tilde\phi_{25} \phi_{54}
\phi_{42}-\phi_{73} \tilde\phi_{36} \phi_{67}-\tilde\phi_{25}
\phi_{57} \phi_{72}+\phi_{57} \phi_{73} \phi_{35} \\ \nonumber 
-\tilde\phi_{35} \phi_{54} \phi_{43}+ \tilde\phi_{35}  \phi_{56}
\phi_{67} \phi_{72} \phi_{13}-\phi_{35} \phi_{56} \phi_{64} \phi_{42}
\phi_{13} .
\eean
Renaming the gauge groups $(1/2,3,4,5,6,7)\rightarrow (4,1,5,3,2,6)$
as well as the fields
$$
\phi_{35}\rightarrow X_{13},~~~~ \tilde \phi_{35}\rightarrow Y_{13},~~~~
\tilde\phi_{36}\rightarrow X_{12},~~~~\phi_{43}\rightarrow X_{51},
$$
we get
\bean
W=X_{12} X_{25} X_{51}+X_{43} X_{35} X_{54}-X_{61} X_{12}
X_{26}-X_{43} X_{36} X_{64}+X_{36} X_{61} X_{13} \\ \nonumber 
-Y_{13} X_{35} X_{51}+Y_{13} X_{32} X_{26} X_{64} X_{41}-X_{13} X_{32}
X_{25} X_{54} X_{41}.
\eean
This is precisely, upto an overall minus sign, the superpotential for
$dP3_{II}$.  Likewise,
the quiver of the $dP3$ model is reproduced exactly, as shown in
\fref{dp4todp3_2}.
\EPSFIGURE[ht]{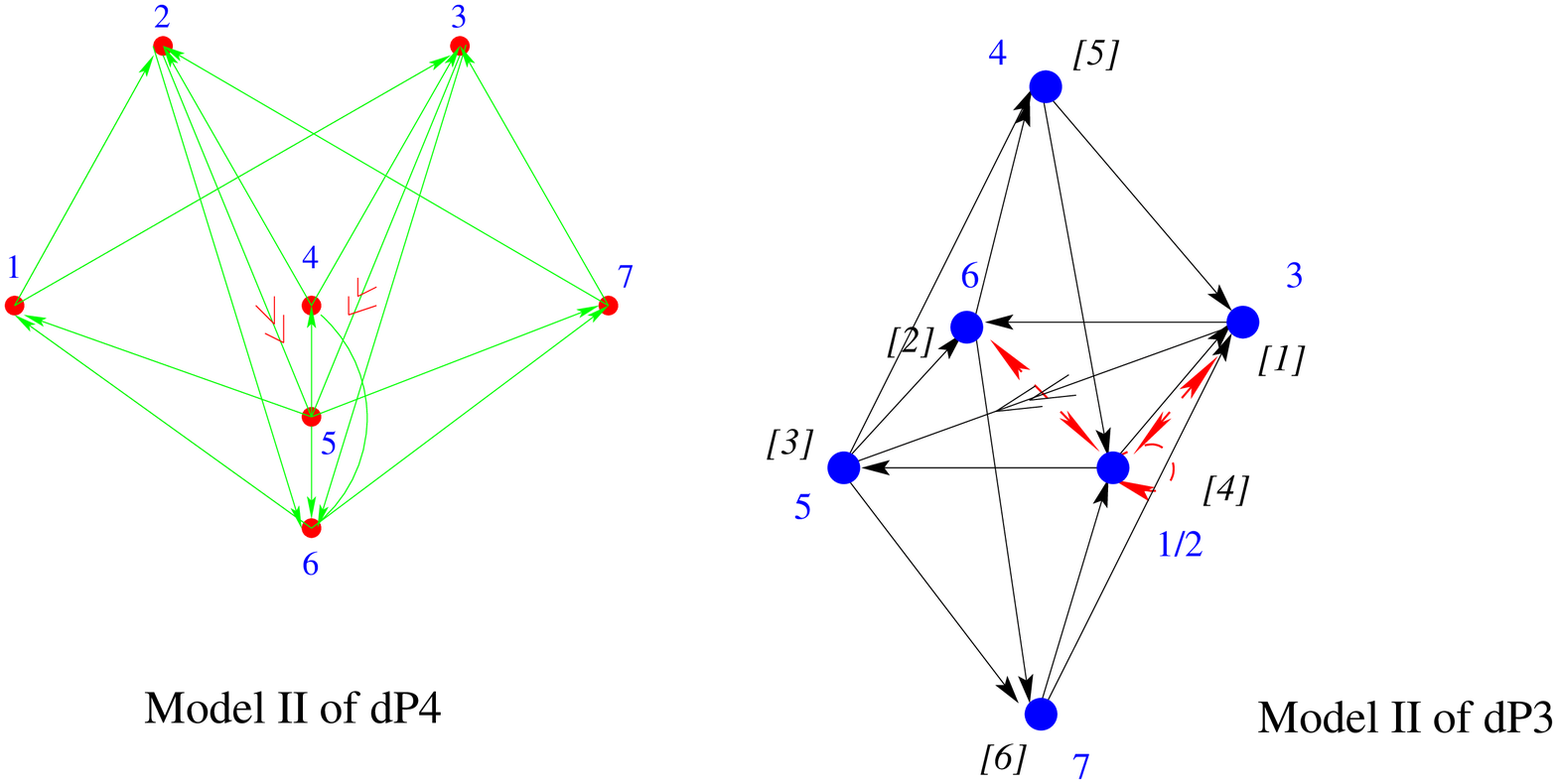,width=15cm}
{Higgsing the field $\phi_{12}$ of $dP4_{II}$ to reach $dP3_{II}$. 
We have used
the dashed lines to indicate the fields to be integrated out and
nodes in bracket to indicate the corresponding nodes in $dP3_{II}$. 
\label{dp4todp3_2}
}

We can get the model II of $dP3$ also from phase III of $dP4$,
whose superpotential is given by
\bean
W=\phi_{62} \phi_{24} \phi_{46}-\phi_{62} \phi_{21}
\phi_{16}+\phi_{34} \phi_{45} \phi_{53}-\phi_{31} \phi_{15}
\phi_{53}+\phi_{57} \phi_{73} \phi_{35}-\phi_{35} \phi_{56} \phi_{63}
\\ \nonumber 
+\phi_{63} \phi_{31} \phi_{16}-\phi_{73} \phi_{34} \phi_{46}
\phi_{67}-\phi_{24} \phi_{45} \phi_{57} \phi_{72}+\phi_{21} \phi_{15}
\phi_{56} \phi_{67} \phi_{72} .
\eean

Setting $\braket{\phi_{73}}=1$, $U(1)_{(7)}\times U(1)_{(3)}$ is
broken down to $U(1)_{[3]}$. During the higgsing, $\phi_{35}$ and $\phi_{57}$
become massive, with equations of motion
\bean
\phi_{35}=\phi_{24} \phi_{45} \phi_{72} \\ \nonumber
\phi_{57}=\phi_{56}\phi_{63}.
\eean

Finally, renaming nodes $(1,2,3/7,4,5,6) \rightarrow (5,2,3,6,4,1)$,
and calling the two fields connecting nodes 1 and 3 in the final
theory 
$$
\phi_{67} \rightarrow \tilde X_{13}, \qquad
\phi_{63} \rightarrow Y_{13},
$$
we get
\bean
W=X_{12} X_{26} X_{61}-X_{12} X_{25}
X_{51}+X_{36} X_{64} X_{43}-X_{35} X_{54}
X_{43}+Y_{13} X_{35} X_{51} \\ \nonumber 
-X_{36} X_{61} X_{13}-X_{26} X_{64} X_{41}
Y_{13} X_{32}+X_{25} X_{54} X_{41} X_{13}
X_{32},
\eean
which is again the superpotential
for phase II of $dP3$.

%
\subsection{Phase III of $dP3$}
This time, we can start from any of the models I, II and III of $dP4$
to reach model III of $dP3$.
First we start from $dP4_{III}$ and turn on a VEV for 
$\phi_{56}$. The fields $\phi_{35}$ and $\tilde\phi_{63}$ will become
massive. Then, in the IR 
we have a $U(1)^6$ theory with 14 fields. Taking
$\braket{\phi_{56}}=1$, the superpotential is 
\bean
W=-\phi_{31} \phi_{15} \tilde\phi_{53}+\phi_{34} \phi_{45}
\tilde\phi_{53}-\tilde\phi_{62} \phi_{21} \phi_{16}+\tilde\phi_{62}
\phi_{24} \phi_{46}- \phi_{24} \phi_{45} \phi_{57} \phi_{72}\\
\nonumber 
+\phi_{57} \phi_{73} \phi_{31} \phi_{16}-\phi_{73} \phi_{34} \phi_{46}
\phi_{67}+\phi_{21} \phi_{15} \phi_{67} \phi_{72} 
\label{higgsing3}
\eean
Let us rename the $U(1)$ gauge factors as $(1,2,3,4,5/6,7)\rightarrow 
(1,4,2,3,5,6)$ and call the fields X, except
$$
\ba{cc}
\phi_{16}\rightarrow X_{15}, & \phi_{15}\rightarrow Y_{15}, \\
\phi_{67}\rightarrow X_{56}, & \phi_{57}\rightarrow Y_{56}, \\ 
\phi_{46}\rightarrow X_{35}, & \phi_{45}\rightarrow Y_{35}.
\ea
$$
Then, after redefining $X_{41} \rightarrow -X_{41}$ and $X_{43} \rightarrow -X_{43}$, 
the superpotential becomes
\bean
W=-X_{21} Y_{15} X_{52}+X_{23} Y_{35} X_{52}+X_{54} X_{41}
X_{15}-X_{54} X_{43} X_{35}+X_{43} Y_{35} Y_{56} X_{64} \\ \nonumber 
+X_{21} X_{15} Y_{56} X_{62}-X_{62} X_{23} X_{35} X_{56}-X_{41} Y_{15}
X_{56} X_{64},
\eean
which is the superpotential for phase III
of $dP3$. The quiver of this model is also correct, as drawn in
\fref{dp4todp3_3}.
\EPSFIGURE[ht]{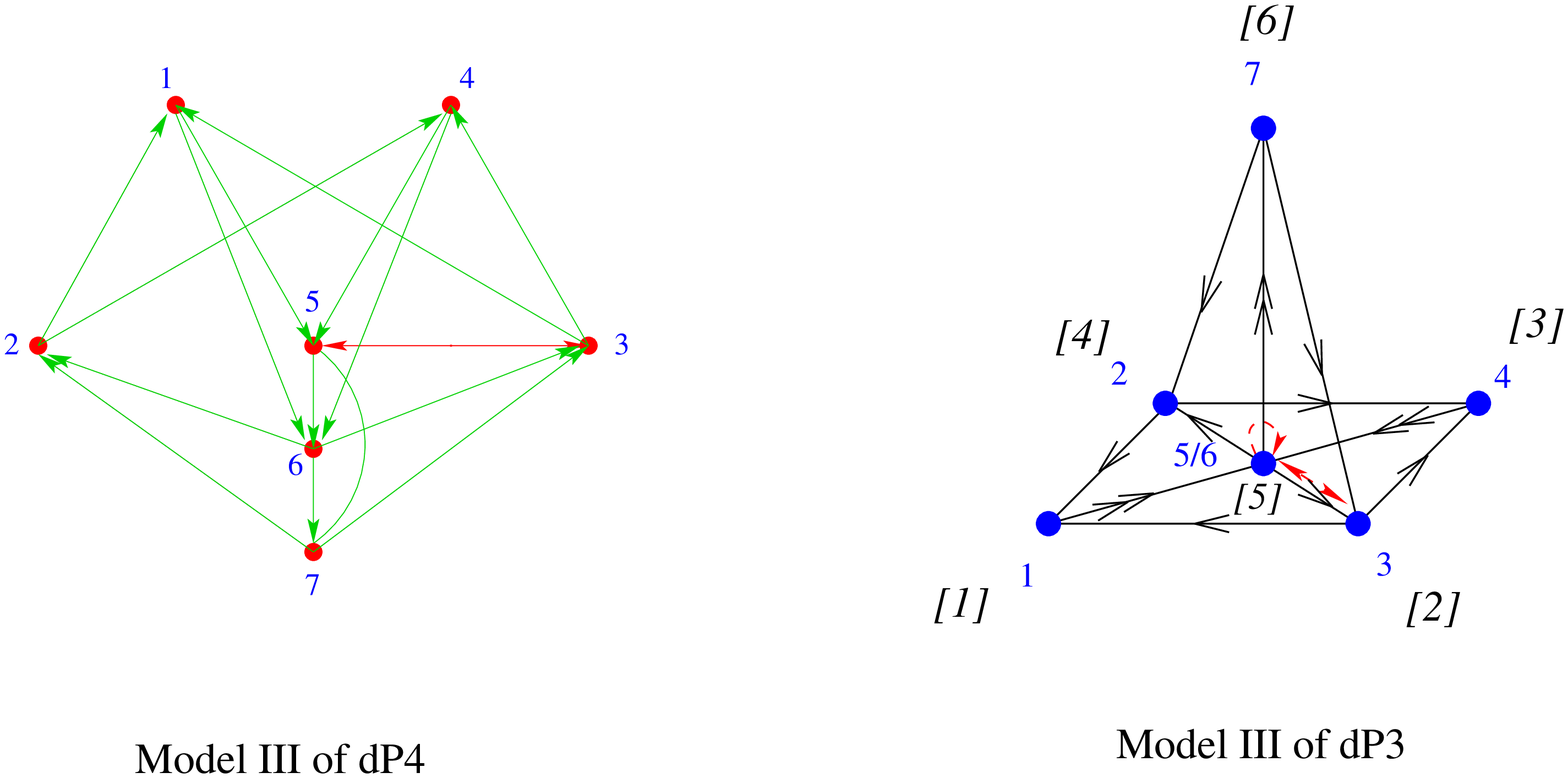,width=15cm}
{Higgsing the field $\phi_{56}$ of $dP4_{III}$ to reach $dP3_{III}$. 
We have used
the dashed lines to indicate the fields to be integrated out and
nodes in bracket to indicate the corresponding nodes in $dP3_{III}$.
\label{dp4todp3_3}
}

Next we start from from phase I of $dP4$, whose superpotential is
\bean
W&=&\phi_{73} \phi_{35} \phi_{57}+\phi_{51} \phi_{12}
\phi_{25}+\phi_{24} \phi_{46} \phi_{61} \phi_{12}-\phi_{73} \phi_{34}
\phi_{46} \phi_{67} \\ \nonumber 
&&-\phi_{45} \phi_{57} \phi_{72} \phi_{24}-\phi_{35} \phi_{56} \phi_{61}
\phi_{13}+\phi_{25} \phi_{56} \phi_{67} \phi_{72}-\phi_{51} \phi_{13}
\phi_{34} \phi_{45}.
\eean

We turn on $\braket{\phi_{56}}=1$. In this case, no mass terms are
generated. After renaming nodes $(1,2,3,4,5/6,7)\rightarrow
(1,4,2,6,5,3)$, calling 
\bean
\phi_{51}\rightarrow X_{51}, & \phi_{61}\rightarrow Y_{51}, & \phi_{67}\rightarrow X_{53} \\
\phi_{57}\rightarrow Y_{53}, & \phi_{46}\rightarrow X_{65}, & \phi_{45}\rightarrow Y_{65},   
\eean
and changing the signs $X_{46} \rightarrow -X_{46}$ and $Y_{65} \rightarrow -Y_{65}$, 
the superpotential becomes
\bean
W&=&X_{32} X_{25} Y_{53}+X_{51} X_{14} X_{45}-X_{25} Y_{51} X_{12}+
X_{45} X_{53} X_{34}-X_{46} X_{65} Y_{51} X_{14} \\ \nonumber 
&&-X_{32} X_{26} X_{65} X_{53}-Y_{65} Y_{53} X_{34} X_{46}+X_{51} X_{12} X_{26} Y_{65}.
\eean
We recognise this to be the superpotential for Phase III of $dP3$
after charge conjugation.

Finally we start from phase II of $dP4$ with superpotential
\bean
W&=&-\phi_{51} \phi_{12} \phi_{25}+\phi_{57} \phi_{73}
\phi_{35}+\phi_{12} \tilde\phi_{26} \phi_{61}-\phi_{73}
\tilde\phi_{36} \phi_{67}+\tilde\phi_{35} \phi_{51}
\phi_{13}-\tilde\phi_{25} \phi_{57} \phi_{72} \\ \nonumber 
&&+\tilde\phi_{25} \phi_{54} \phi_{42}-\phi_{35} \phi_{54}
\phi_{43}-\tilde\phi_{26} \phi_{64} \phi_{42}+\tilde\phi_{36}
\phi_{64} \phi_{43}+\phi_{25} \phi_{56} \phi_{67} \phi_{72}-\phi_{35}
\phi_{56} \phi_{61} \phi_{13}.
\eean

Setting $\braket{\phi_{64}}=1$, $U(1)_{(6)}\times U(1)_{(4)}$ breaks to
the $U(1)_{[3]}$ and mass terms are generated for
$\tilde\phi_{26}$, $\phi_{42}$, $\tilde\phi_{36}$ and
$\phi_{43}$, with equations of motion
\bean
\tilde\phi_{26}=\tilde\phi_{25} \phi_{54} & \ \ \ \ \ &
\phi_{42}=\phi_{12} \phi_{61} \\ 
\tilde\phi_{36}=\tilde\phi_{35} \phi_{54} & & \phi_{43}=\phi_{73}
\phi_{67} \ .
\eean

Relabelling nodes $(1,2,3,4/6,5,7) \rightarrow (4,1,3,6,5,2)$, changing
$X_{15} \rightarrow -X_{15}$ and calling 
$$ 
\ba{cc}
\phi_{25} \rightarrow X_{15}, & \tilde \phi_{25} \rightarrow Y_{15} \\
\phi_{35} \rightarrow Y_{35}, & \tilde \phi_{35} \rightarrow X_{35} \\
\phi_{54} \rightarrow X_{56}, & \phi_{56} \rightarrow Y_{56},
\ea
$$
we get
\bean
W&=&X_{54} X_{41} X_{15}+X_{52} X_{23} Y_{35}+X_{35} X_{54} X_{43}
-Y_{15} X_{52} X_{21}-X_{15} Y_{56} X_{62} X_{21} \\ \nonumber 
&&-Y_{35} Y_{56} X_{64} X_{43}-X_{23} X_{35} X_{56} X_{62}+Y_{15} X_{56} X_{64} X_{41},
\eean
and once again obtain the theory for $dP3_{III}$.
%
%


\subsection{Phase IV of $dP3$}

After the above detailed demonstrations, we will be brief in this
part. In this case, we start from the model II of $dP4$ and give the
nonzero VEV to $\phi_{56}$. It is easy to see the quiver will be that
of model IV of
$dP3$, as drawn in \fref{dp4todp3_4}. Renaming nodes $(1,2,3,4,5/6,7)
\rightarrow (2,4,5,1,6,3)$ 
and making the following replacements
\bean
\phi_{51} \rightarrow X_{62},~~~~~~& \phi_{61} \rightarrow Y_{62},~~~~~~& \\
\phi_{25} \rightarrow Z_{46},~~~~~~& \widetilde{\phi}_{25} \rightarrow Y_{46},
       ~~~~~~&\widetilde{\phi}_{26} \rightarrow  X_{46}, \\
\phi_{57} \rightarrow X_{63},~~~~~~& \phi_{67} \rightarrow Y_{63},~~~~~~ & \\
\phi_{35} \rightarrow Z_{56},~~~~~~& \widetilde{\phi}_{35} \rightarrow X_{56},
       ~~~~~~&\widetilde{\phi}_{36} \rightarrow  Y_{56}, \\
\phi_{54} \rightarrow Y_{61},~~~~~~& \phi_{64} \rightarrow X_{61}.~~~~~~&
\eean
we get the correct superpotential, upto an overall minus sign and the charge conjugation
of all fields suggested by the condensed quiver.

\EPSFIGURE[ht]{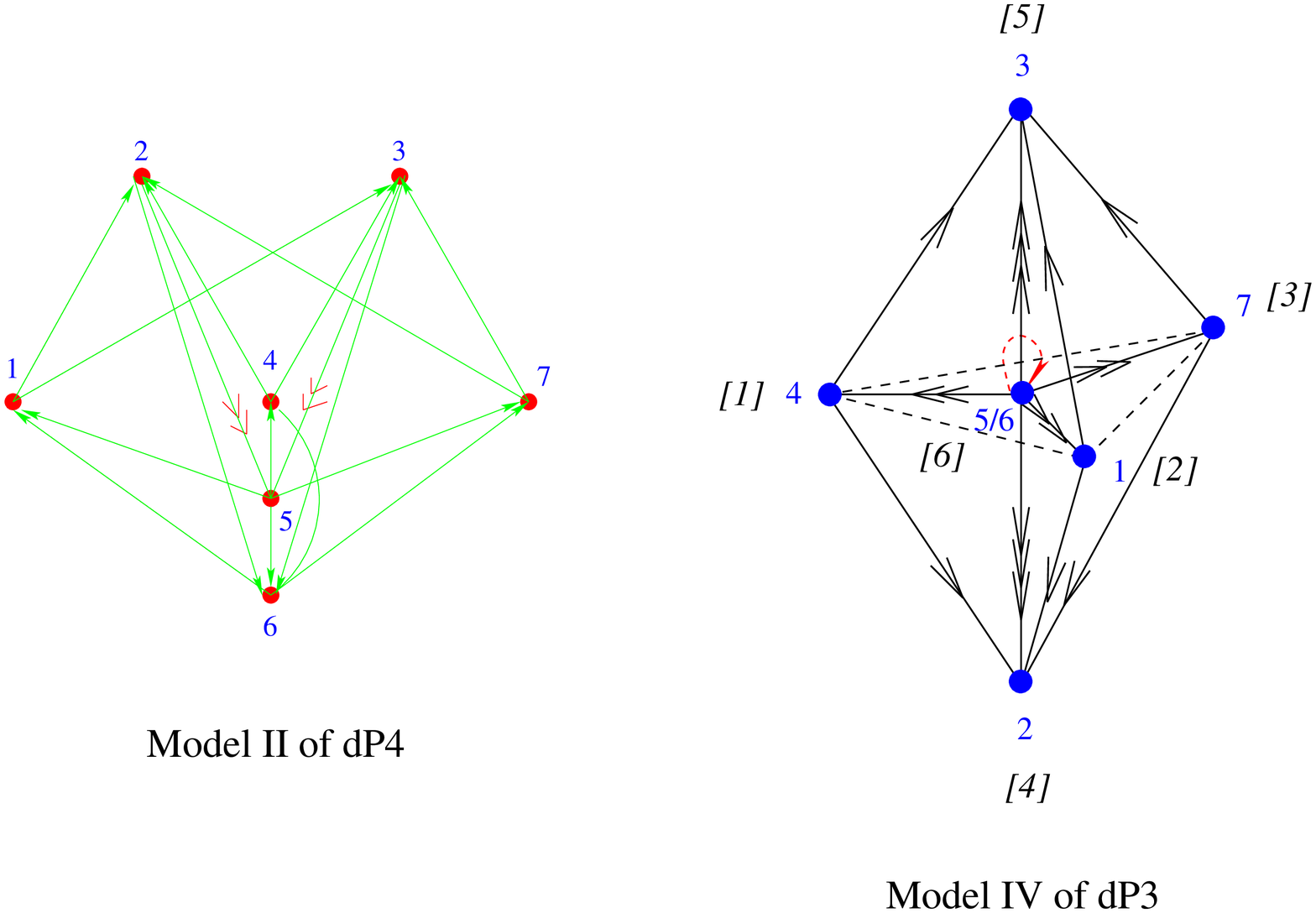,width=15cm}
{Higgsing the field $\phi_{56}$ from $dP4_{II}$ to give $dP3_{IV}$. 
We have used the dashed lines to indicate the fields to be integrated out and
nodes in bracket to indicate the corresponding nodes in model IV of $dP3$.
\label{dp4todp3_4}
}
%
\section{PdP4: del Pezzo Three Blownup at a Non-Generic Point}
\label{sec:toric}
We have obtained, via the unhiggsing method, three toric phases of a new
theory from the four phases of the cone over del Pezzo 3. 
In the
previous sections, because we have used the quivers obtained from
$(p,q)$-web techniques in,
\cite{Hanany:2001py}, we have assumed that we have arrived at the
theory for $dP4$.
Is this indeed so? The purpose of this section is to show
that we are not quite right, even though we did arrive at a new theory
which is $dP3$ blownup at one point.

Let us begin with the model $dP4_{III}$ obtained from unhiggsing.
We recall the matter content and superpotential here:
\bean
W_{III} & = & [\phi_{62} \phi_{24} \phi_{46}-\phi_{62}
  \phi_{21} \phi_{16}]+[\phi_{34} \phi_{45} \phi_{53} -\phi_{31} 
 \phi_{15} \phi_{53}] \\
& + & \phi_{57} \phi_{73} \phi_{35} -\phi_{35} \phi_{56} \phi_{63} +
  \phi_{63} \phi_{31} \phi_{16} \nonumber \\
& - & \phi_{73} \phi_{34} \phi_{46} \phi_{67} -\phi_{24}\phi_{45} \phi_{57}
  \phi_{72} +  \phi_{21} \phi_{15}  \phi_{56}  \phi_{67}\phi_{72}
\eean
and
\[
d_{III} =
\mat{\phi_{15} & \phi_{16} & \phi_{21} & \phi_{24} & \phi_{31} & \phi_{3
   4} & \phi_{35} & \phi_{45} & \phi_{46} & \phi_{53} & \phi_{56} & \phi_{5
   7} & \phi_{62} & \phi_{63} & \phi_{67} & \phi_{72} & \phi_{73} \cr 
    -1 & 
    -1 & 1 & 0 & 1 & 0 & 0 & 0 & 0 & 0 & 0 & 0 & 0 & 0 & 
   0 & 0 & 0 \cr 0 & 0 & -1 & 
    -1 & 0 & 0 & 0 & 0 & 0 & 0 & 0 & 0 & 1 & 0 & 0 & 1 & 
   0 \cr 0 & 0 & 0 & 0 & -1 & -1 & 
    -1 & 0 & 0 & 1 & 0 & 0 & 0 & 1 & 0 & 0 & 1 \cr 0 & 0 & 
   0 & 1 & 0 & 1 & 0 & -1 & 
    -1 & 0 & 0 & 0 & 0 & 0 & 0 & 0 & 0 \cr 1 & 0 & 0 & 0 & 
   0 & 0 & 1 & 1 & 0 & -1 & -1 & 
    -1 & 0 & 0 & 0 & 0 & 0 \cr 0 & 1 & 0 & 0 & 0 & 0 & 0 & 
   0 & 1 & 0 & 1 & 0 & -1 & -1 & 
    -1 & 0 & 0 \cr 0 & 0 & 0 & 0 & 0 & 0 & 0 & 0 & 0 & 0 & 
   0 & 1 & 0 & 0 & 1 & -1 & -1 \cr}.
\]

We can obtain all the 17 F-terms from $W_{III}$:
\[
\begin{array}{l}
\phi_{21} \phi_{56} \phi_{67} \phi_{72} = \phi_{31} \phi_{53}, \quad
  \phi_{31} \phi_{63} = \phi_{21} \phi_{62}, \quad
  \phi_{15} \phi_{56} \phi_{67} \phi_{72} = \phi_{16} \phi_{62}, \quad
  \phi_{46} \phi_{62} = \phi_{45} \phi_{57} \phi_{72}\\
  \phi_{16} \phi_{63} = \phi_{15} \phi_{53}, \quad
  \phi_{45} \phi_{53} = \phi_{46} \phi_{67} \phi_{73}, \quad
  \phi_{57} \phi_{73} = \phi_{56} \phi_{63}, \quad
  \phi_{34} \phi_{53} = \phi_{24} \phi_{57} \phi_{72}, \quad\\
  \phi_{24} \phi_{62} = \phi_{34} \phi_{67} \phi_{73}, \quad
  \phi_{34} \phi_{45} = \phi_{15} \phi_{31}, \quad
  \phi_{15} \phi_{21} \phi_{67} \phi_{72} = \phi_{35} \phi_{63}, \quad
  \phi_{35} \phi_{73} = \phi_{24} \phi_{45} \phi_{72}, \quad\\
  \phi_{24} \phi_{46} = \phi_{16} \phi_{21}, \quad
  \phi_{16} \phi_{31} = \phi_{35} \phi_{56}, \quad
  \phi_{15} \phi_{21} \phi_{56} \phi_{72} = \phi_{34} \phi_{46} \phi_{73}, \quad
  \phi_{15} \phi_{21} \phi_{56} \phi_{67} = \phi_{24} \phi_{45}
  \phi_{57},\\ 
  \phi_{35} \phi_{57} = \phi_{34} \phi_{46} \phi_{67}.
\end{array}
\]

These are all monomial relations! These F-terms thus generate a toric
ideal. This is suggestive that our moduli space is actually toric
and thus cannot be the cone over the generic del Pezzo
4. Let us perform the Forward Algorithm of \cite{toric} to check.

From the F-terms we can actually express the solution space in terms
of the $K$-matrix prescribing a cone:
\[
K^T = 
\tmat{
 & \phi_{15} & \phi_{16} & \phi_{21} & \phi_{24} & \phi_{31} &
 \phi_{34} 
& \phi_{35} & \phi_{45} & \phi_{46} & \phi_{53} & \phi_{56} &
 \phi_{57} & \phi_{62} & \phi_{63} & \phi_{67} & \phi_{72} & \phi_{73}
 \cr  \phi_{15} & 1 & 2 & 0 & 1 & -1 & 0 & 1 & 0 & 1 & 1 & 0 & 0 & 
    -1 & 0 & 0 & 0 & 0 \cr \phi_{2
   1} & 0 & 1 & 1 & 1 & 0 & 0 & 1 & 0 & 1 & 1 & 0 & 0 & 
    -1 & 0 & 0 & 0 & 0 \cr \phi_{34} & 0 & 
    -1 & 0 & 0 & 1 & 1 & 0 & 0 & -1 & 
    -1 & 0 & 0 & 1 & 0 & 0 & 0 & 0 \cr \phi_{45} & 0 & -1 & 0 & 
    -1 & 1 & 0 & 0 & 1 & 0 & -1 & 0 & 0 & 1 & 0 & 0 & 0 & 0 \cr 
\phi_{56} & 0 & 2 & 0 & 1 & 0 & 0 & 1 & 0 & 1 & 1 & 1 & 0 & -1 & 
    -1 & 0 & 0 & 0 \cr \phi_{57} & 0 & -1 & 0 & -1 & 0 & 0 & 
    -1 & 0 & 0 & 0 & 0 & 1 & 1 & 1 & 0 & 0 & 0 \cr \phi_{6
   7} & 0 & 1 & 0 & 1 & 0 & 0 & 1 & 0 & 0 & 1 & 0 & 0 & 0 & 0 & 1 & 
   0 & 0 \cr 
\phi_{72} & 0 & 1 & 0 & 0 & 0 & 0 & 1 & 0 & 1 & 1 & 0 & 0 & 0 & 0 & 0 & 
   1 & 0 \cr \phi_{73} & 0 & -1 & 0 & 0 & 0 & 0 & -1 & 0 & 
    -1 & 0 & 0 & 0 & 1 & 1 & 0 & 0 & 1 \cr
},
\]
where we express the 17 variables in terms of 9 as we read from the
above vertically: $\phi_{j=1,\ldots,17} = \prod_{i=1}^9
\phi_i^{K^T_{ij}}$.

Equipped with the $d$ and $K$ matrices we can now easily perform the
Forward Algorithm to obtain the moduli space as a toric variety. The
answer is:
\[
G_t= \mat{
0 & 0 & 0 & 0 & 1 & 1 & 1 & 1 & 1 & 1 & 1 & 1 & 1 & 1 & 1 & 1 & 2 & 2 \cr 
    -1 & 0 & 0 & 1 & -1 & -1 & 0 & 0 & 0 & 0 & 0 & 0 & 0 & 0 & 0 & 1 & 
    -1 & 0 \cr 2 & 1 & 1 & 0 & 1 & 1 & 0 & 0 & 0 & 0 & 0 & 0 & 0 & 0 & 0 & 
    -1 & 0 & -1 }.
\]

We immediately see that after a permutation $s$ and an $SL(3;\IZ)$
transformation, which certainly do not effect the moduli space, we can
bring the above $G_t$ to a familiar form:
\[
\begin{array}{l}
\mat{0 & 0 & 0 & 1 & 1 & 1 & 2 & 2 \cr -1 & 0 & 1 & -1 & 0 & 1 & 
    -1 & 0 \cr 2 & 1 & 0 & 1 & 0 & -1 & 0 & -1 \cr}
\stackrel{s}{\Rightarrow}
\mat{0 & 1 & 2 & 0 & 1 & 2 & 0 & 1 \cr 1 & 0 & -1 & 0 & -1 & 0 & 
    -1 & 1 \cr 0 & 0 & 0 & 1 & 1 & -1 & 2 & -1 \cr
}\\
\stackrel{\tmat{0 & 0 & 1 \cr 0 & -1 & -1 \cr 1 & 2 & 1
\cr}}{\Longrightarrow}
\mat{
0 & 0 & 0 & 1 & 1 & -1 & 2 & -1 \cr -1 & 0 & 1 & -1 & 0 & 1 & 
    -1 & 0 \cr 2 & 1 & 0 & 1 & 0 & 1 & 0 & 2 \cr
}
\end{array}
\]

We recognise the embedding of this toric diagram into our familiar
orbifold $\IC^3/(\IZ_3 \times \IZ_3)$ in \fref{f:z3z3embed}. We have
explicitly labelled the multiplicity of the GLSM fields (homogeneous
coordinates) and see that it is perfectly congruent with the
observations in \cite{multiplicity,Muto} about the emergence of the
Pascal's triangle.
\EPSFIGURE[h]{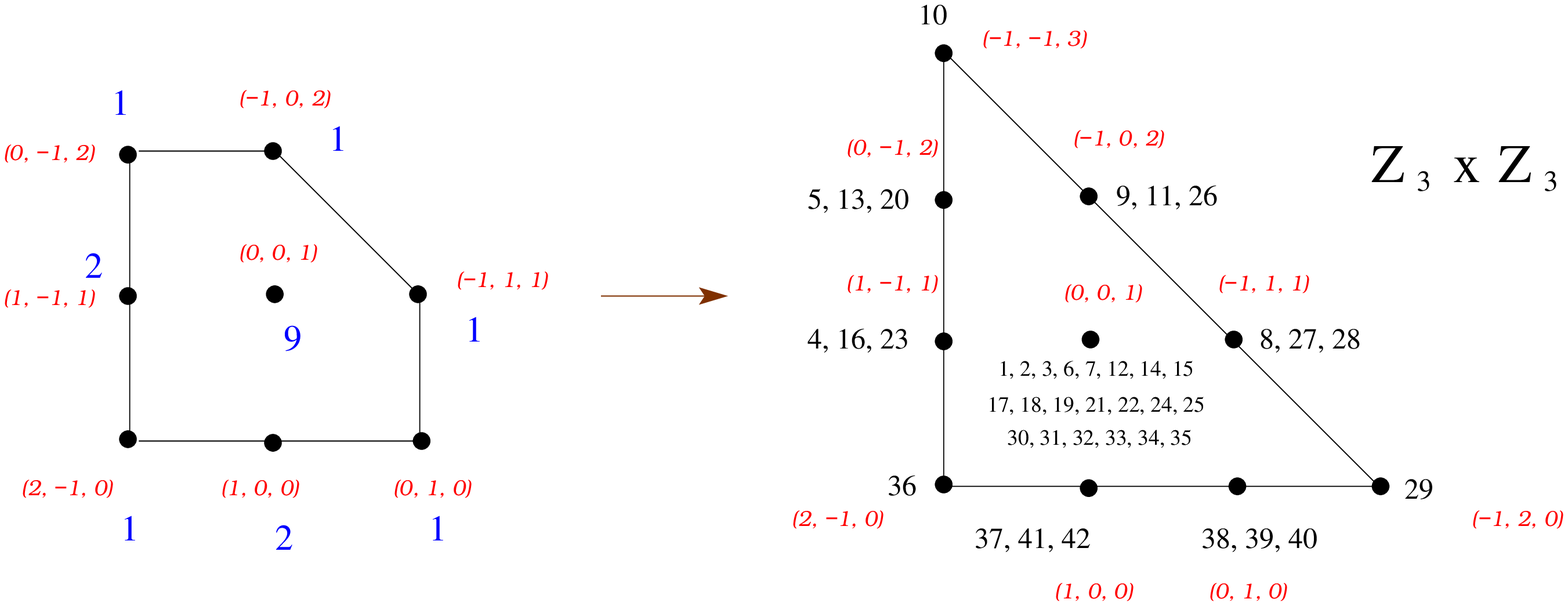,width=15cm}
{The embedding of the moduli space for the model III obtained from
unhiggsing $dP3$, into the toric diagram of $\IC^3/(\IZ_3 \times
\IZ_3)$. We have labelled all coordinates explicitly. In the left
the numbers (in blue) are the multiplicities corresponding to the
nodes (q.v.~\cite{multiplicity,Muto}) and in the right, the numbers are the
GLSM fields commonly used.
\label{f:z3z3embed}
}

What we have for the moduli space is therefore a toric variety which
is a blowup of $dP3$.
According to \cite{Leung:1997tw}, such a cone is no longer over
an ample surface. Therefore whatever theory we have obtained, is not
that of the generic del Pezzo 4 theory, because all del Pezzo surfaces
are ample; to this point we shall return in the next subsection.

Certainly, unhiggsing the $dP3$ corresponds to blowing it up at a
point and the so-called $dP4$ theories in the previous sections are
indeed $dP3$ blowup at a point and hence the cone over
$\IP^2$ blowup at 4
points. We thus conclude that the theories we have obtained in the
previous sections must be the cone over
$\IP^2$ blownup at 4 {\bf non-generic}
points. We shall henceforth call this variety
the {\bf Pseudo dP4}, or PdP4.

Here is an important fact: whereas $\IP^2$ blownup at {\bf
  generic} points are the del Pezzo surfaces, as we shall see below, 
blowing up at {\em non-generic} points no longer gives
us del Pezzo surfaces in the strict sense. Indeed as remarked above,
our $dP4$ is really a toric variety while the generic del Pezzo $k$ for $k
\ge 4$ certainly is not.
Recently such non-generic del Pezzos have
risen in the context of \cite{loops}.
\subsection{Some Properties of the Moduli Space}
\label{subsec:dp4}
We here have a toric variety whose toric diagram is given in
\fref{f:z3z3embed}; let us determine some of its geometrical
properties in light of the discussion above that it should be a
$PdP4$.

Let us study the compact surface as a projective variety because we
know the properties of the del Pezzo surfaces well; our $dP4$ is
simply an affine cone thereover. In other words we shall study the
so-called Pseudo del Pezzo surface $PB4$ in comparison to the true
$B4$.

First, given the toric diagram, one could immediately find the
characteristic classes using combinatorics \cite{Fulton}. With the
convenient help of the package Macaulay \cite{Mac}, 
we immediately arrive at the Betti numbers:
$b^0 = b^4 = 1, b^1 = b^3 = 0$ and $b^2 = 5$. Indeed the
middle-dimensional homology of $\IP^2$ blown up at 4 points
should consist of the
hyperplane $\IP^1$ class as well as 4 exceptional divisors of the
blowup. Thus we pass our preliminary test of homology.

Next let us study the explicit embedding as projective varieties. 
We know, using the method
of fat points\footnote{YHH would like to thank Hal Schenck of Texas
	AMU for extensive discussions on this point.} 
in $\IP^2$ \cite{fat} that the generic del Pezzo 4 surface can be
embedded as the smooth intersection of 5 quadrics in
$\IP^5$ (q.v.~\cite{Hartshorne}).
The affine cone over it, would have an isolated singular point at the
conical apex (say, at the origin) and it is this point upon which we
place our D-brane probe.
A non-generic one however, say 3 co-linear points being
blown-up, may have more complicated presentation. 
Moreover, the precise
positions of the blowups determine the complex structure moduli space
of the $B_4$, whereupon singularities may arise as one varies these
positions and causes the Jacobian matrix to be non-maximal rank. In
these cases the affine cone $dP4$ would have singularities at 
more than the point at the origin. To these we refer as
non-generic, or pseudo $dP4$'s.

From the toric diagram in \fref{f:z3z3embed}, we can instantly
determine the projective embedding by finding the relations of the
homogeneous coordinate ring \cite{Cox}.
We find that we obtain the
intersection of 5 quadrics in $\IP^5$; and indeed upon computing the
Jacobian matrix of the variety we find non-trivial singular loci.
Therefore our toric diagram
corresponds to a non-generic $dP4$, or $PdP4$.

Let us re-iterate to our reader that a surface given by a toric
diagram of the form \fref{f:z3z3embed} does not have ample
anti-canonical class (and hence not del Pezzo).
Standard results from
toric geometry (e.g., \cite{Fulton}) dictates that
a Cartier divisor $D$ on a complete toric
variety $X$ is ample iff its support function is strictly
convex. Combinatorially this translates to the following:
Let $X$ be a complete toric variety with fan $\Sigma = \{ \sigma_a \}$
with each cone $\sigma$ generated by $\{v_i \}$ as $\sigma= \sum_i
\IR_{\ge 0} v_i$. A divisor $D$ can be written as $\sum_{i=1}^r a_i
D_i$ with $D_i$ corresponding to $v_i$, then
\begin{theorem}
$D$ is ample iff for each cone $\sigma$ there exists an integer vector
$m_\sigma$ such that $\langle m_\sigma, v_i \rangle = -a_i$ for all $i$
and such that $m_\sigma \ne m_\tau$ for different cones $\sigma$ and $\tau$.
\label{ample}
\end{theorem}

The anticanonical class is of course given by $K = - \sum_{i=1}^r D_i$
with all $a_i = -1$. We can thus easily proceed with the check for
ampleness. The surface we have at hand has the fan
as given in \fref{f:z3z3embed}: 
$\Sigma = \{ \sigma_{i=1 \ldots 7} \}$ with
$\sigma_i$ generated by $\{v_i, v_{(i+1) \bmod 7}\}$ where $\{ v_1, v_2, \ldots
v_7 \}$ $=$ $\{ \tmat{1 \cr 0}, \tmat{0 \cr 1}, \tmat{-1 \cr 1},
\tmat{-1 \cr 0}, \tmat{-1 \cr -1}, \tmat{0 \cr -1}, \tmat{1 \cr -1}
\}$.
The list of support function $m_\sigma$ can be easily computed as
$\{\tmat{1 \cr 1}, \tmat{0 \cr 1}, 
\tmat{-1 \cr 0}, \tmat{-1 \cr 0}, \tmat{0 \cr -1}, \tmat{-2 \cr -1},
\tmat{1 \cr 2} \}$. Due to the repetition therein we conclude that
$-K$ is indeed not ample and our surface is not del Pezzo.

In fact all the toric diagrams which satisfy the criteria of the above
theorem are classified in dimension 2 \cite{Oda} and are precisely the
del Pezzo polytopes; \fref{f:z3z3embed} is certainly not a member of
the classification.

\subsection{Confirmations from the Inverse Algorithm}

Having assertained that the moduli space is actually toric with the
explicit toric diagram and embedding given in \fref{f:z3z3embed},
we can naturally use the conjecture that toric duality is Seiberg duality
\cite{seiberg,Beasley:2001zp} to see whether we indeed obtain the
above phases. We will use the algorithm of the multiplicity symmetry
introduced in \cite{multiplicity}.

We have 3 models which we must obtain. Starting from the 42 GLSM fields of
$\IZ_3 \times \IZ_3$ in \fref{f:z3z3embed}, we obtain a total of 216
theories which fall into various isoclasses.
If we keep, for example, the fields $\{$ 4, 5, 6, 7, 8, 9, 12, 14, 15, 
18, 21, 22, 23, 30, 36, 37, 38, 42 $\}$, we obtain the theory with
17 fields, precisely the model III addressed above.
If, on the other hand, we kept the fields $\{$ 4, 5, 6, 7, 8, 11, 12,
17, 18, 19,
21, 22, 23, 24, 25, 30, 32, 36, 37, 38, 42 $\}$, then the resulting
theory is the model II with 19 fields.

These consistency checks are very re-assuring: even though the moduli space
we obtained is not that of the cone over the generic del Pezzo four, 
it is a perfectly
well-defined toric Calabi-Yau variety. Most importantly, toric duality
from the Inverse Algorithm indeed reproduces the Seiberg dual theories
obtained from field theoretic analyses using unhiggsing.

However we have yet to obtain the model I with 15 fields. This poses a hitherto
unencountered problem. The Inverse Algorithm does not give us any theories with
15 fields. What seems to be wrong?
Let us attempt to find the moduli space of Model I using the Forward
Algorithm. From the superpotential 
\bean
 W &=& - X_{12}\, X_{25}\,  X_{51} +
     X_{13}\,  X_{34}\,  X_{45}\,  X_{51} +
     X_{12}\,  X_{24}\,  X_{46}\,  X_{61} -
     X_{13}\,  X_{35}\,  X_{56}\,  X_{61} - \\
&&   X_{24}\,  X_{45}\,  X_{57}\,  X_{72} +
     X_{25}\,  X_{56}\,  X_{67}\,  X_{72} +
     X_{35}\,  X_{57}\,  X_{73} -
     X_{34}\,  X_{46}\,  X_{67}\,  X_{73},
\eean
we can solve for the 15 F-terms as
\beq
\label{ftermsdP4I}
\begin{array}{ccc}
 \{  X_{12}, X_{73}, X_{35}, X_{57},
    X_{25}, X_{51}\} &=& \pm
 \{ \sqrt{\frac{ X_{13}\, X_{34}\, X_{45}\,
       X_{56}\, X_{67}\, X_{72}}{ X_{24}\,
       X_{46}\, X_{61}}},
    \sqrt{\frac{ X_{13}\, X_{24}\, X_{45}\, X_{56}\,
       X_{61}\, X_{72}}{ X_{34}\, X_{46}\,
       X_{67}}},
	\sqrt{\frac{ X_{24}\, X_{34}\,
       X_{45}\, X_{46}\, X_{67}\, X_{72}}
      { X_{13}\, X_{56}\, X_{61}}},\\
&& \sqrt{\frac{ X_{13}\, X_{34}\, X_{46}\, X_{56}\,
       X_{61}\, X_{67}}{ X_{24}\, X_{45}\,
       X_{72}}},
	\sqrt{\frac{ X_{13}\, X_{24}\,
       X_{34}\, X_{45}\, X_{46}\, X_{61}}
      { X_{56}\, X_{67}\, X_{72}}},
    \sqrt{\frac{ X_{24}\, X_{46}\, X_{56}\, X_{61}\,
       X_{67}\, X_{72}}{ X_{13}\, X_{34}\,
       X_{45}}}\}.
\end{array}
\eeq

These are not monomial relations! In fact no attempt of the solution space
(the so-called space of commuting matrices \cite{Chris1}) of these
F-terms could give purely monomial relations. In other words, we cannot
generate a $K$-matrix which corresponds to an
integral polyhedral cone. The Forward Algorithm
thus already fails to be valid.

This is rather surprising. We have checked in Subsection
\ref{subsec:dp4} that the moduli space is toric and in particular, the
toric $PdP4$. Furthermore we have
checked above that we indeed consistently obtain models $dP4_{II}$ and
$dP4_{III}$. Indeed we must be able to obtain this remaining model of
$dP4_I$ from partial resolutions.
Yet, the Forward Algorithm (and thus necessarily
the Inverse Algorithm) already does not seem to succeed 
to generate a cone, and hence a toric description.

The situation however, is easily remedied. The F-terms in
\eref{ftermsdP4I} generate a cone over $\IQ$ instead of our usual
circumstance of $\IZ$. It corresponds to a K-matrix with $\frac12$
entries due to the square root exponent; we simply reconvert our basis
and work in a large integral cone by multiplying it by 2:
$$
\ba{ccc}
{\tiny \mat{ \frac{1}{2} & 1 & 0 & \frac{1}{2} & 0 & -
     \frac{1}{2}   & 0 & 0 & - \frac{1}{2}
       & 0 & \frac{1}{2} & 0 & 0 & 0 & \frac{1}
   {2} \cr - \frac{1}{2}   & 0 & 1 & \frac{1}
   {2} & 0 & \frac{1}{2} & 0 & 0 & \frac{1}{2} & 0 & -
     \frac{1}{2}   & 0 & 0 & 0 & \frac{1}
   {2} \cr \frac{1}{2} & 0 & 0 & \frac{1}{2} & 1 & \frac{1}
   {2} & 0 & 0 & - \frac{1}{2}   & 0 & \frac{1}
   {2} & 0 & 0 & 0 & - \frac{1}{2}   \cr 
    \frac{1}{2} & 0 & 0 & \frac{1}{2} & 0 & \frac{1}
   {2} & 1 & 0 & - \frac{1}{2}   & 0 & -
     \frac{1}{2}   & 0 & 0 & 0 & \frac{1}
   {2} \cr - \frac{1}{2}   & 0 & 0 & \frac{1}
   {2} & 0 & \frac{1}{2} & 0 & 1 & \frac{1}{2} & 0 & 
    \frac{1}{2} & 0 & 0 & 0 & - \frac{1}{2} 
      \cr \frac{1}{2} & 0 & 0 & - \frac{1}{2} 
      & 0 & - \frac{1}{2}   & 0 & 0 & \frac{1}
   {2} & 1 & \frac{1}{2} & 0 & 0 & 0 & \frac{1}{2} \cr 
    - \frac{1}{2}   & 0 & 0 & \frac{1}
   {2} & 0 & - \frac{1}{2}   & 0 & 0 & \frac{1}
   {2} & 0 & \frac{1}{2} & 1 & 0 & 0 & \frac{1}{2} \cr 
    \frac{1}{2} & 0 & 0 & - \frac{1}{2} 
      & 0 & \frac{1}{2} & 0 & 0 & \frac{1}{2} & 0 & \frac{1}
   {2} & 0 & 1 & 0 & - \frac{1}{2}   \cr 
    \frac{1}{2} & 0 & 0 & - \frac{1}{2} 
      & 0 & \frac{1}{2} & 0 & 0 & \frac{1}{2} & 0 & -
     \frac{1}{2}   & 0 & 0 & 1 & \frac{1}{2} \cr  }}
&
\stackrel{\times 2}{\Longrightarrow}
&
K^T = 
\tmat{
1 & 2 & 0 & 1 & 0 & -1 & 0 & 0 & 
    -1 & 0 & 1 & 0 & 0 & 0 & 1 \cr 
    -1 & 0 & 2 & 1 & 0 & 1 & 0 & 0 & 1 & 0 & 
    -1 & 0 & 0 & 0 & 1 \cr 1 & 0 & 0 & 1 & 2 & 1 & 0 & 0 &
   -1 & 0 & 1 & 0 & 0 & 0 & 
    -1 \cr 1 & 0 & 0 & 1 & 0 & 1 & 2 & 0 & -1 & 0 & 
    -1 & 0 & 0 & 0 & 1 \cr 
    -1 & 0 & 0 & 1 & 0 & 1 & 0 & 2 & 1 & 0 & 1 & 0 & 0 & 
   0 & -1 \cr 1 & 0 & 0 & -1 & 0 & 
    -1 & 0 & 0 & 1 & 2 & 1 & 0 & 0 & 0 & 1 \cr 
    -1 & 0 & 0 & 1 & 0 & 
    -1 & 0 & 0 & 1 & 0 & 1 & 2 & 0 & 0 & 1 \cr 1 & 0 & 0 &
   -1 & 0 & 1 & 0 & 0 & 1 & 0 & 1 & 0 & 2 & 0 & 
    -1 \cr 1 & 0 & 0 & -1 & 0 & 1 & 0 & 0 & 1 & 0 & 
    -1 & 0 & 0 & 2 & 1 \cr
}
\ea
$$
Now application of the standard Forward Algorithm on this integral
matrix K and the incidence matrix for the quiver in \fref{f:dP4I}
readily gives us (after an appropriate unimodular transformation that
does not change the geometry) the correct toric diagram in the left of
\fref{f:z3z3embed}.

Therefore with the caveat of needing to convert a rational cone to an
integral one, upon which both the Forward and Inverse Algorithms
depend, we have shown that the remaining case of $dP4_I$ also gives
the same toric variety. Therefore all 3 Seiberg dual phases for $dP4$
give the same moduli space, as expected. More importantly, the moduli
space is toric, an affine variety which is
a cone over $\IP^2$ blown up at 4 colinear points. Thus in our
notation, the models $dP4_{I,II,III}$ should really be called
$PdP4_{I,II,III}$ and to this convention we shall henceforth adhere.
\section{Unhiggsing $PdP4$ Once Again to $PdP5$}
Having obtained the toric, non-generic, $PdP4$, it is natural
to ask whether this pattern should continue. In other words, could we
unhiggs/blowup this non-generic $dP4$ to something else that is
perhaps also toric, and in particular, $PdP5$?

We shall see that this indeed is the case in this section, whereby
confirming out unhiggsing procedure as well as the Inverse
Algorithm. We find that
there are in fact four toric phases
which are related to each other by Seiberg duality. Without much ado
let us present the results below.
\subsection{Model $PdP5_I$}
\EPSFIGURE[ht]{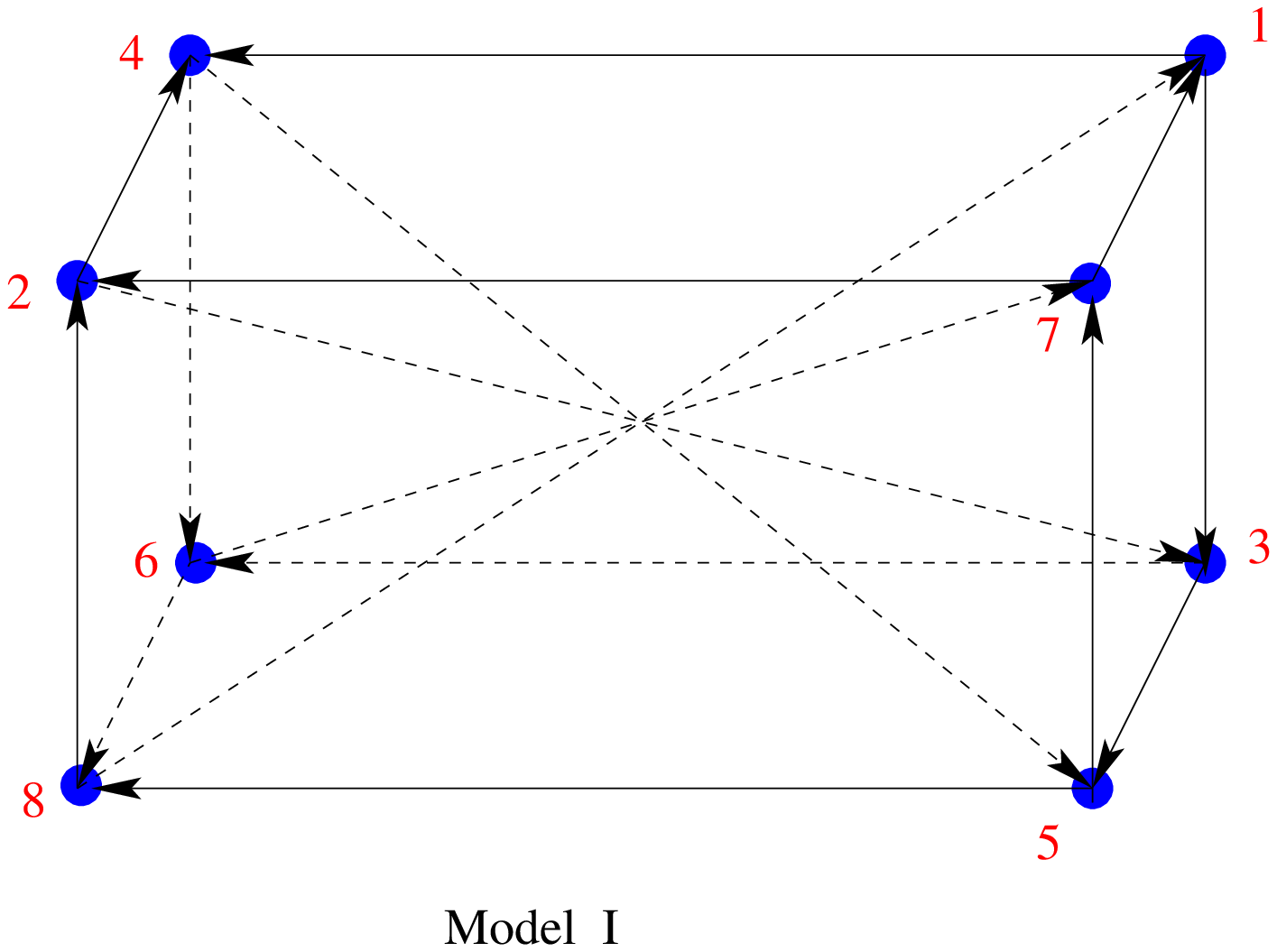,width=8cm}
{Model I of the $PdP5$ theory, unhiggsed from $PdP4_I$.
\label{dP5_1}
}
Now we unhiggs the above $PdP4$ to the $dP5$ given in \cite{Hanany:2001py}.
Indeed it will turn out that it is not really $dP5$ either and we will use
the notation $PdP5$ for pseudo-$dP5$. Comparing the quiver of model I of
$PdP5$ in \cite{Hanany:2001py} with the quiver of $PdP4_I$, we 
see that giving the field $\phi_{68}$ nonzero VEV is the way to higgs
$PdP5$ to $PdP4$. 

Since the $PdP5_I$ has 16 fields while $PdP4_I$ has
15, there is no mass term generated
in the higgsing process and the unhiggsing is straight-forward. We just
need to lift the superpotential of $PdP4_I$ directly. 
With a little of algebra we reach the superpotential \eref{W_dP5_1}:
\begin{eqnarray}
\label{W_dP5_1}
W_I=\phi_{13} \phi_{35} \phi_{58} \phi_{81}+\phi_{14} \phi_{46}
\phi_{68} \phi_{81} + \phi_{35} \phi_{57} \phi_{72} \phi_{23}-\phi_{46}
\phi_{67} \phi_{72} \phi_{24} \\ \nonumber 
+\phi_{67} \phi_{71} \phi_{13} \phi_{36}-\phi_{57} \phi_{71} \phi_{14}
\phi_{45}+\phi_{58} \phi_{82} \phi_{24} \phi_{45}-\phi_{68} \phi_{82}
\phi_{23} \phi_{36} .
\end{eqnarray}

Let us analyze the symmetry of the quiver. First there is a cyclic
$\IZ_4$ symmetry around the horizontal axis. Second, there is a $\IZ_2$
symmetry which exchanges $(1357)$ and $(2864)$ and reverses the arrows
(i.e., charge
conjugation). The superpotential preserves both symmetries. It is easy
to see that by redefining the signs of 
fields $\phi_{58},\phi_{13},\phi_{46},\phi_{14},\phi_{23}$ we can
regroup the superpotential as
\bean
W_I & = & [(1358)-(3572)+(5714)-(7136)]+\\
&&[(3682)-(5824)+(7246)-(1468)],
\eean
where the four terms in brackets are related to each other by $\IZ_4$
and two brackets are related to each other by $\IZ_2$.

Due to the abundance of such
symmetries, Seiberg dualising any of the nodes
will give the same result. We will
call, without loss of generality, the result from dualising on node 8,
$PdP5_{II}$, upon which we shall in the ensuing subsections
continue to dualise to obtain models $PdP5_{III,IV}$.
\subsection{Model $PdP5_{II}$}
Without loss of generality, let us dualise $PdP5_I$ on node 8. 
Since there are no cubic terms in \eref{W_dP5_1}, no mass terms are
generated. The resulting 
model has 20 fields, with no bi-directional arrows. 
It is then possible to call all fields $\phi_{ij}$, and the
superpotential is 
\EPSFIGURE[ht]{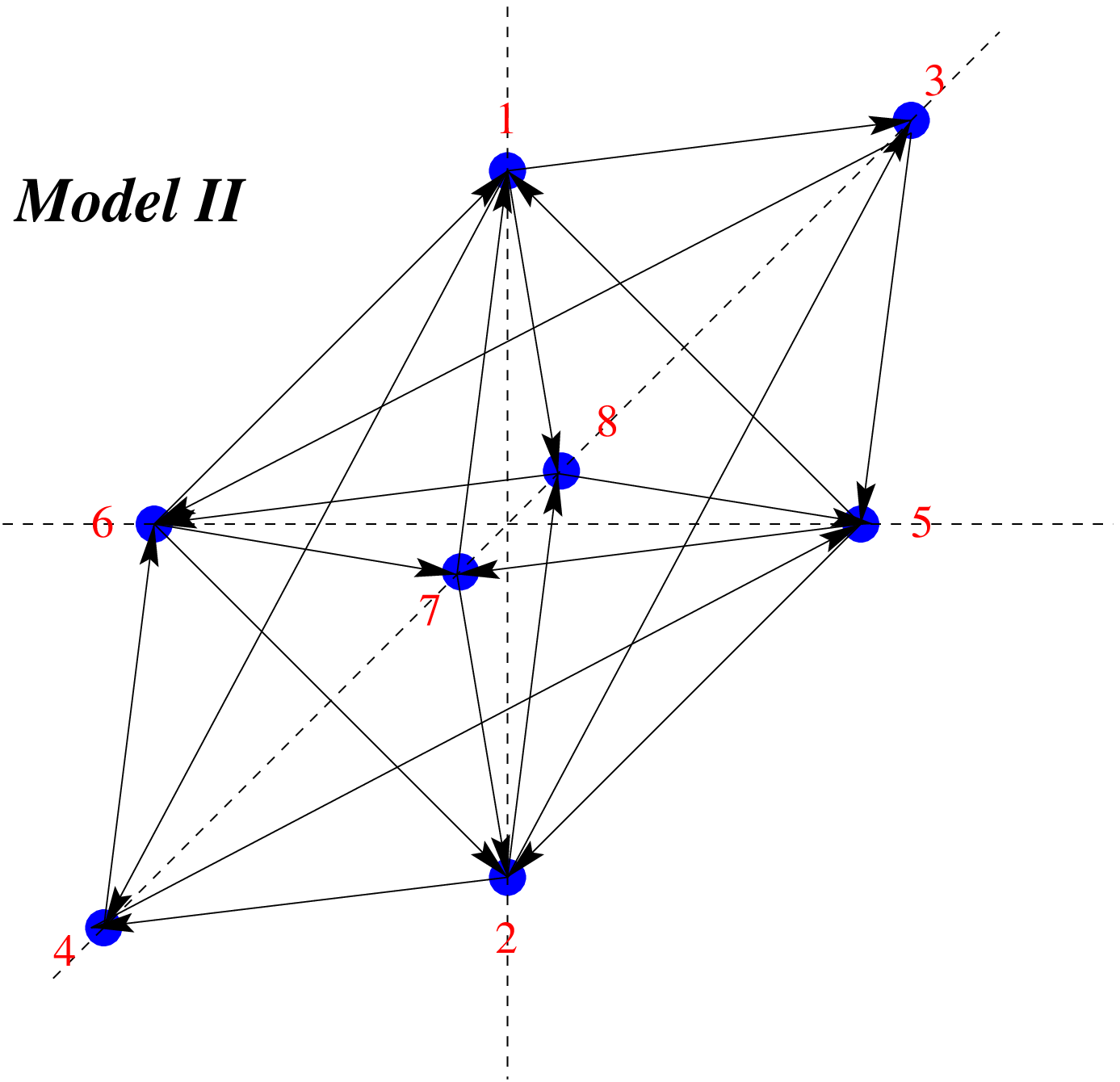,width=6.5cm}
{The quiver for the theory $PdP5_{II}$. This is a nice 
3D representation, with nodes $7$ and $8$ located at the centre.
\label{dP5_2}
}
\begin{eqnarray}
\label{W_dP5_2}
W=\phi_{13} \phi_{35} \phi_{51}+\phi_{14} \phi_{46}
\phi_{61}+\phi_{52} \phi_{24} \phi_{45}- \\ \nonumber
\phi_{62} \phi_{23}
\phi_{36}-\phi_{51} \phi_{18} \phi_{85}  
-\phi_{61} \phi_{18} \phi_{86}- \\ \nonumber
\phi_{52} \phi_{28} \phi_{85}+\phi_{62}
\phi_{28} \phi_{86}+\phi_{35} \phi_{57} \phi_{72} \phi_{23} - \\
\nonumber 
\phi_{46} \phi_{67} \phi_{72} \phi_{24}+\phi_{67} \phi_{71} \phi_{13}
\phi_{36}-\phi_{57} \phi_{71} \phi_{14} \phi_{45}
\end{eqnarray}
The symmetries are $\IZ_2:(1,5)\leftrightarrow (2,6)$ and 
$\IZ_2:(1,3)\leftrightarrow (2,4)$. Using these we can group the 
superpotential in the orbits of the global symmetries as
\bean
W_{II} & = & [ (7246)+(7145)-(7136)-(7235)]\\
&& +[(513)+(623)-(524)-(614)] \\
 & & +[(528)+(618)-(518)-(628)].
\eean
Geometrically, we let 
$(12)$ be the $z$-axis, $(34)$, the $y$-axis, $(56)$, the $x$-axis and 
$(7,8)$, around the origin as in \fref{dP5_2}; 
Then symmetries are just the rotation with
$x,y,z$ axis. 

From these symmetries, we see that to get new Seiberg dual phase, we
have only two 
choices: dualise node $3$ or node $7$. Starting from node $3$ we 
obtain a new theory: $PdP5_{III}$ and starting from node $7$ we obtain
$PdP5_{IV}$. To these we now turn.

\newpage
\subsection{Model $PdP5_{III}$}
Recall from the above that dualising $PdP5_{II}$ on node 3 we obtain
the quiver in \fref{dP5_3} with
the dual superpotential
\bean
W=M_{15} \phi_{51}+\phi_{14} \phi_{46} \phi_{61}+\phi_{25} \phi_{24}
\phi_{45}-\phi_{62} M_{26}-\phi_{51} \phi_{18} \phi_{85}-\phi_{61}
\phi_{18} \phi_{86} \\ \nonumber 
-\phi_{52} \phi_{28} \phi_{85}+\phi_{62} \phi_{28} \phi_{86}+M_{25}
\phi_{57} \phi_{72}-\phi_{46} \phi_{67} \phi_{72} \phi_{24}+\phi_{67}
\phi_{71} M_{16} \\ \nonumber 
-\phi_{57} \phi_{71} \phi_{14} \phi_{45}-M_{15} \tilde\phi_{53}
\tilde\phi_{31}-M_{16} \tilde\phi_{63} \tilde\phi_{31}-M_{25}
\tilde\phi_{53} \tilde\phi_{32}+M_{26} \tilde\phi_{63} \tilde\phi_{32}.
\eean

We see that $M_{51}$, $\phi_{15}$, $M_{62}$ and $\phi_{26}$ become
massive, leading to a theory with 20 fields. Integrating them out
using their equations of motion (and calling all fields  
$\phi_{ij}$) we finally have
\begin{eqnarray}
W&=&\phi_{14} \phi_{46} \phi_{61}+\phi_{52} \phi_{24}
\phi_{45}-\phi_{61} \phi_{18} \phi_{86}-\phi_{52} \phi_{28} \phi_{85}
+\phi_{25} \phi_{57} \phi_{72}+\phi_{67} \phi_{71} \phi_{17}
\\ \nonumber 
&&-\phi_{16}
\phi_{63} \phi_{31}-\phi_{25} \phi_{53} \phi_{32}
-\phi_{53} \phi_{31} \phi_{18} \phi_{85}+\phi_{63} \phi_{32} \phi_{28}
\phi_{86}-\phi_{46} \phi_{67} \phi_{72} \phi_{24}- 
\phi_{57} \phi_{71} \phi_{14} \phi_{15}
\label{W_dP5_3}
\end{eqnarray}

%
\EPSFIGURE[ht]{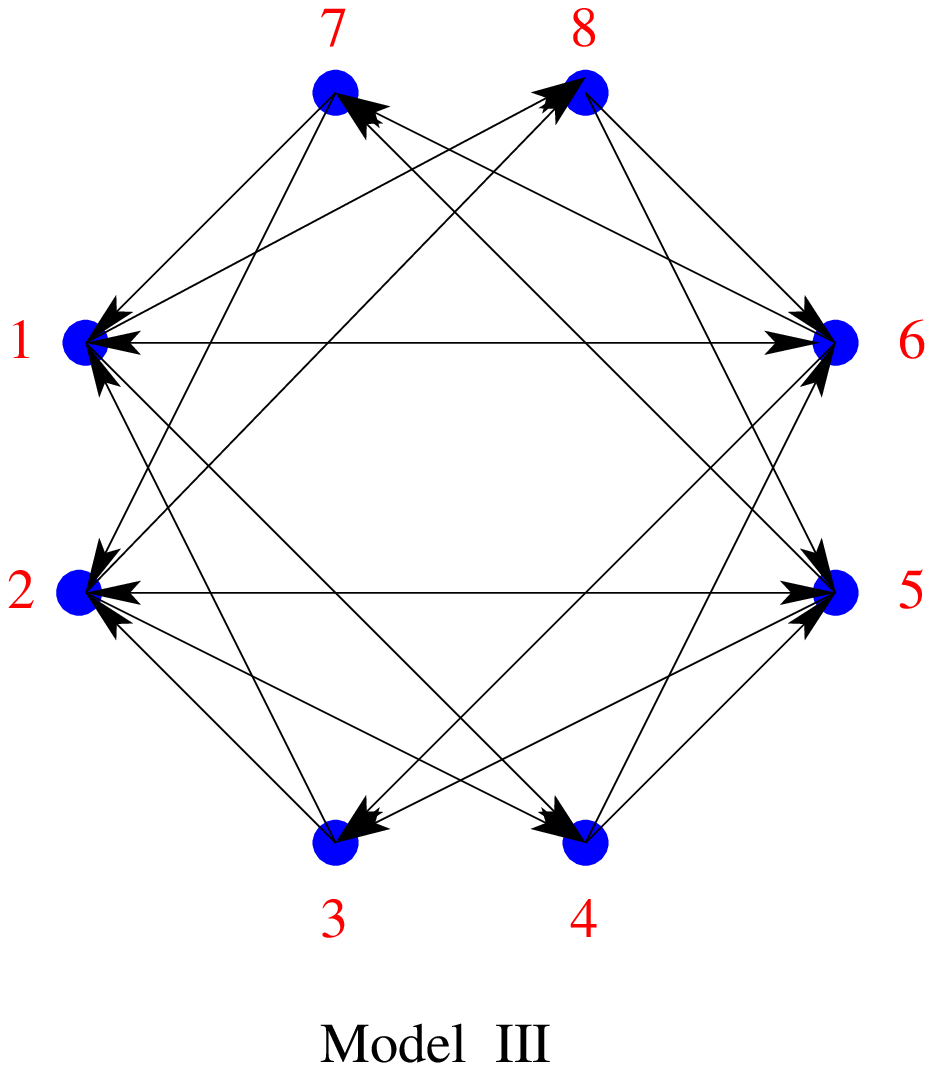,width=8cm}
{The theory for $PdP5_{III}$. Notice the bidirectional arrows (16) and
(25).
\label{dP5_3}
}
From the quiver in \fref{dP5_3} we see that the theory has these
symmetries:
$\IZ_2^{(1)}: (15)\leftrightarrow (26)$,
$\IZ_2^{2)}: (34)\leftrightarrow (78)$ and 
$\IZ_2^{(3)}: (1237)\leftrightarrow (6548)$.
Grouping terms together with respect to this symmetry we get the
superpotential 
\bean
W_{III} & = & [(7145)+(7246)-(3185)-
\\
&&(3286)]+[(528)+(618)-(524)-(614)]
\\ 
& & +[(163)+(253)-(167)-(257)],
\eean
where every bracket is invariant under $\IZ_2^{(1)}\times \IZ_2^{2)}$ while
the first bracket is invariant under $\IZ_2^{(3)}$ and last two brackets
are related by $\IZ_2^{(3)}$.
From the symmetry, we see that no new phase can be reached by Seiberg
duality that still remains toric.
\subsection{Model $PdP5_{IV}$}
Recall that we have a final model which comes from $PdP5_{II}$ after
dualising on node 7. Now this node does not appear in any cubic term
of \eref{W_dP5_2}, thus there are no massive fields. This phase has 24
fields, with the quiver shown in \fref{dP5_4} and the superpotential is
\bean
W=\phi_{13} \phi_{35} \phi_{51}+\phi_{14} \phi_{46}
\phi_{61}+\phi_{52} \phi_{24} \phi_{45}-\phi_{62} \phi_{23} \phi_{36}
-\phi_{51} \phi_{18} \phi_{85}-\phi_{61} \phi_{18} \phi_{86}-\phi_{52}
\phi_{28} \phi_{85}-\phi_{62} \phi_{28} \phi_{86} \\
+\phi_{35} \tilde\phi_{52} \phi_{23}-\phi_{46} \tilde\phi_{62}
\phi_{24}+\tilde\phi_{61} \phi_{13} \phi_{36}-\tilde\phi_{51}
\phi_{14} \phi_{45}
+\tilde\phi_{51} \phi_{17} \phi_{75}-\tilde\phi_{52} \phi_{27}
\phi_{75}-\tilde\phi_{61} \phi_{17} \phi_{76}+\tilde\phi_{62}
\phi_{27} \phi_{76},
\eean
where we have indicated the Seiberg mesons with tildes.

The theory has the symmetry: $\IZ_2^{(1)}:(15)\leftrightarrow (26)$, 
$\IZ_2^{(2)}: (73)\leftrightarrow (84)$ and changing tildes to
non-tildes,  $\IZ_2^{(3)}:(78)\leftrightarrow (43)$ and 
$\IZ_2^{(4)}: (173)\leftrightarrow (284)$ (here the
tildes are not changed to non-tilde).

\EPSFIGURE[ht]{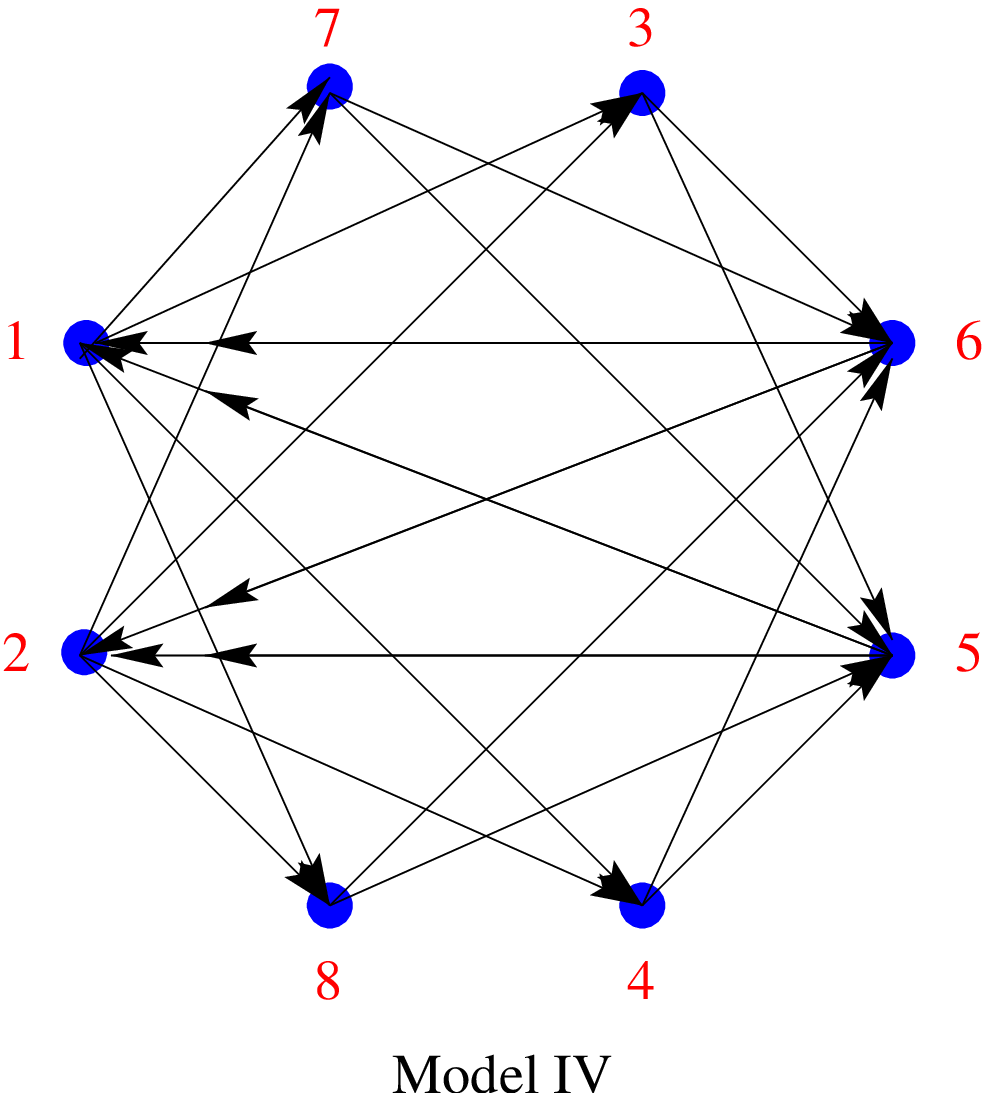,width=8cm}
{The quiver diagram for model $PdP5_{IV}$.
\label{dP5_4}
}

The superpotential can be accordingly grouped as
\bean
W_{IV} & = & \{ - \ [(\widetilde{62}7)+(\widetilde{51}7)+(628)+(518)] + \\
&&	[(\widetilde{62}4)+(\widetilde{51}4)+(623)+(513)] \} + \\
&&	\{ [(\widetilde{61}7)+(\widetilde{52}7)+(618)+(528)] - \\
&&	[(614)+(524)+(\widetilde{61}3)+(\widetilde{52}3)] \}.
\eean

Here every bracket is grouped by $\IZ_2^{(1)}\times \IZ_2^{(2)}$; the
first and second brackets as well as the third and the fourth are
each grouped by $\IZ_2^{(3)}$. Moreover these two pairs of brackets
([1][2]) and ([3][4])
are related to each other by $\IZ_2^{(4)}$. 
From the symmetry, we see that nodes $3,4,7,8$
are equivalent to each other, so Seiberg duality can not give new
phase that is toric.

In conclusion then, by unhiggsing the $PdP4$ theory we have obtained a
toric phase for a blownup of thereof, which we have called $PdP5$. By
applying Seiberg duality, we have found all toric phases
of this theory and there are 4 of these: $PdP5_{I,II,III,IV}$. 
We summarize them in \fref{f:dP5all}.

\EPSFIGURE[ht]{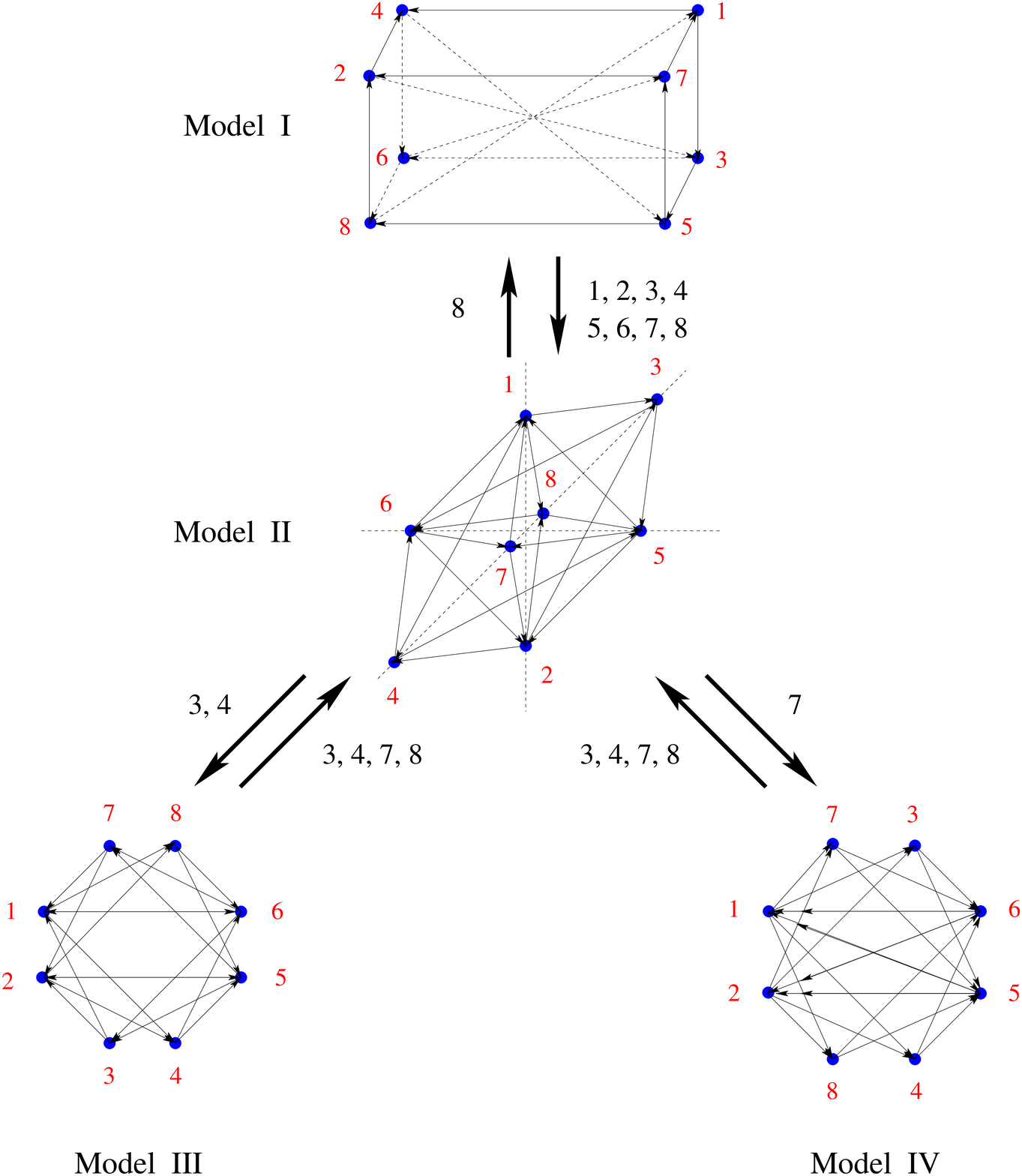,width=15cm}
{The quivers for the four Seiberg dual phases of $dP5$ which all have
rank one nodes. The nodes upon which one dualises to transform between
them are shown next to the arrows.
\label{f:dP5all}
}
%
%
\subsection{PdP5 and the Orbifolded Conifold}

In the vein of thought of Section \ref{sec:toric}, let us investigate
whether continuing this method of blowing up/unhiggsing would give
rise to yet another toric moduli space, in particular the cone over
the so-called pseudo del Pezzo 5.

Let us study model $dP5_I$; the others are related thereto by Seiberg
duality and hence have the same moduli space.  
From the quiver from \fref{dP5_1}
and the superpotential in \eref{W_dP5_1}, we can readily proceed with
the Forward Algorithm of \cite{DGM,Aspinwall,DD,Chris1,toric}.
The final moduli space we obtain is summarised in the $G_t$ matrix for
the toric diagram:
$$
G_t = \tmat{
    0 & 0 & 0 & 0 & 1 & 1 & 1 & 1 & 1 & 1 & 1 & 1 & 1 & 1 & 1 & 1 & 1 & 1 & 
    1 & 1 & 2 & 2 & 2 & 2 \cr -1 & 0 & 0 & 1 & -1 & 
     -1 & 0 & 0 & 0 & 0 & 0 & 0 & 0 & 0 & 0 & 0 & 0 & 0 & 1 & 1 & 
     -1 & 0 & 0 & 1 \cr 2 & 1 & 1 & 0 & 1 & 1 & 0 & 0 & 0 & 0 & 0 & 0 & 0 & 
    0 & 0 & 0 & 0 & 0 & -1 & -1 & 0 & -1 & -1 & -2 \cr  }.
$$
This corresponds to the toric diagram as shown in \fref{f:dP5toric}
(we have performed the usual $SL(3; \IZ)$ transformation so as to make
the presentation compatible with the standard $\IZ_3 \times \IZ_3$
toric diagram in \cite{Chris1,toric}).

\EPSFIGURE[ht]{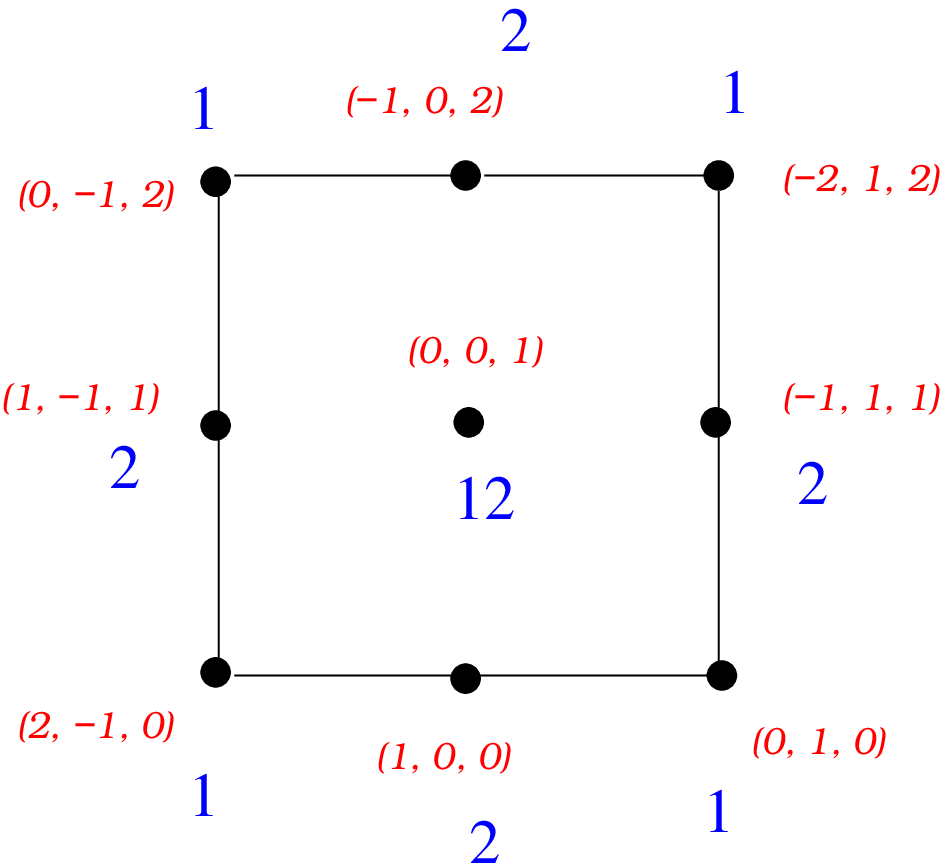,width=8cm}
{Toric diagram for the theory corresponding to $dP3$ blownup twice, or
what have called the toric $PdP5$. We have shown the coordinates in
red and the multiplicity of the GLSM fields, in blue.
\label{f:dP5toric}
}

The toric diagram is indeed expected, consistent with the
node-addition of Section \ref{sec:toric}. Incidentally, the
multiplicities of the GLSM fields (shown in blue next to the nodes)
are still consistent with the observations of \cite{multiplicity,Muto}.
What might be a surprise to the reader is that this is a well-known
toric diagram (q.v.~e.g.~\cite{Karch,Oh,Uranga2}) and the
superpotential has been known from the brane diamond techniques.

\fref{f:dP5toric} is the orbifolded conifold, with the affine equation
$$
xy = z^2 \quad uv = z^2.
$$
Therefore the surface over which our Calabi-Yau is an affine cone
is a compact divisor (4-cycle)
in the orbifolded conifold $C_{2,2}$.

Along the lines of Subsection \ref{subsec:dp4}, let us again check the
geometry we have obtained. As before, we focus on the compact
projective surface over which our moduli space is an affine
cone. Computing the homology of the toric surface corresponding to
\fref{f:dP5toric}, we obtain $b^0 = b^4 = 1, b^1 = b^3 = 0$ and $b^2 =
6$. This is indeed the homology of $\IP^2$ blown up at 5 points.

We proceed to check the embedding equations. We recall that
$\IP^2$ blown up at 5 generic points is the well-known del Pezzo
surface of degree 4, as the intersection of 2 quadrics in $\IP^4$.  
If we have say, 5 non-generic points however, we could again use \cite{fat}
to find that such a surface is given by 2 quadrics in $\IP^4$, but
with non-trivial singular loci.
On the other hand, the homogeneous coordinate ring of the toric
variety in \fref{f:dP5toric} gives us
precisely such an embedding into $\IP^4$. Thus we have shown that
our moduli space is indeed a toric variety, the
non-generic $PdP5$. Once again, we see that checking against Theorem
\ref{ample}, the anticanonical divisor is not ample and our surface is
not del Pezzo.

Therefore with the current technology of the inherently toric method
of $(p,q)$ webs which provided us the quivers, the unhigssing
procedure stays within the toric realm.  The unhiggsing can bring us
from $\IP^2$ to del Pezzo 3, and continue so to the surfaces
corresponding to $\IP^2$ blownup at 4 and 5 special points, which
we have rather cavalierly called the non-generic or pseudo del Pezzos.
We summarise our results for the unhiggsing/blowups in
\fref{f:blowups}.
\section{Quiver Symmetries and the Superpotential}

In this section, we will try, in the spirit of
\cite{multiplicity,soliton},
to use symmetry arguments to fix the superpotentials for the theories
which we have called $PdP4_{I,II}$ in Section \ref{sec:unhiggs}.
The situation is more complex here than the cases discussed 
in \cite{multiplicity} because
$PdP4_{III}$ does not have any explicit symmetry. More precisely, the
quiver has some symmetry but the superpotential breaks it. We will
show here, by certain consistency arguments, that we can sometimes 
determine how the superpotential breaks the quiver symmetry.
\subsection{Symmetries of $PdP4_I$}
Let us start from model I. We recall from \fref{f:dP4I} that there is
an explicit
$\IZ_2^{(1)} \times \IZ_2^{(2)}$ 
($1\leftrightarrow 7$ and $2\leftrightarrow 3$)
quiver symmetry; here we list the orbits of loops (i.e., possible terms in
the superpotential) under this group\footnote{In fact, there is another 
set of gauge invariant operators ${(1257346),(7251346),(1357246),(7351246)}$
which contains the seven nodes. It is easy to concluded that this orbit has 
to be excluded, so we will neglect it in the following discussion.}:
\begin{enumerate}
\item $\{ (125), (725), (135), (735)   \}$ 
\item $\{ (1256), (7256), (1356), (7356) \}$ 
\item $\{ (1245),(7245), (1345), (7345) \}$ 
\item $\{ (1246), (7246), (1346), (7346) \}$ 
\item $\{ (12456), (72456), (13456), (73456) \}$.
\end{enumerate}

Now there is yet another group $\IZ_2^{(3)}$ defined by the
simultaneous action of $(4,1,2)\leftrightarrow (6,7,3)$
plus charge conjugation. Under this third $\IZ_2$,  
orbits (1), (4) and (5) are self-dual while 
orbit (2) maps to (3). Therefore under this full
$\IZ_2 \times \IZ_2 \times \IZ_2$ quiver symmetry we have four orbits:
(1), (4), (5) and (2-3).
We can easily count the number of times the fields appear in 
these orbits to be respectively 12, 16, 20 and 32. Now in Section
\ref{sec:toric}, we have assertained that $PdP4_I$ is toric, thus
since it has 15 fields, a total of $15 \times 2 = 30$ fields must
appear in the superpotential \cite{toric,multiplicity}. 
This is incompatible with the orbit counting above. 
Therefore {\it the quiver 
symmetry must be broken.} But how?

First we assume only one $\IZ_2$ is broken. We have the following cases:
(A) $\IZ_2^{(3)}$ is broken, giving us 5 orbits
with number of fields $12,16,16,16$ and $20$. Again it does not work; 
(B) $\IZ_2^{(2)}$ is broken. Now although $2\leftrightarrow 3$ is
broken, by $\IZ_2^{(3)}$ we still have the loop 
$ 3\leftrightarrow 1\leftrightarrow  7\leftrightarrow 2$,  so 
we still have orbits with field numbers $12,16,32, 20$ and it still
can not be;
(C) $\IZ_2^{(1)}$ is broken, giving us the same situation as case (B).
These cases tell us that we must break a combination
of $\IZ_2$'s and leave the diagonal term invariant. It is obvious that 
$\IZ_2^{(3)}$ can not combine with $\IZ_2^{(1)}$ or $\IZ_2^{(2)}$.
This leave us with the only choice (D) breaking the combination 
$\IZ_2^{(3)}$ and $\IZ_2^{(4)}$, defined by the action
$(1,2)\leftrightarrow (7,3) $. This will turn out to be the right
choice.

Now let us write down the orbits of loops under the symmetry 
$\IZ_2^{(3)} \times \IZ_2^{(4)}$:
\begin{itemize}
\item (Ia). $\{ (125), (735) \}$ 
\item (Ib). $\{ (135), (725)   \}$
\item (Ic). $\{ (1246), (7346) \}$ 
\item (Id). $\{  (7246), (1346)\}$
\item (Ie). $\{ (1256),  (7356), (7345), (1245) \}$
\item (If). $\{ (7256), (1356), (7245), (1345)\}$
\item (Ig). $\{ (12456), (73456) \}$
\item (Ih). $\{ (72456), (13456)\}$
\end{itemize}
The number of fields in the orbits are respectively $6,6,8,8,16,16,
10$ and 10.
There are these ways to get the number 30:
$6+8+16, 6+6+8 +10$. 

For the choice of $6+6+8+10$, From  
orbits (Ia) and (Ib), we  get $(125)-(135)+(735)-(725)$ where
we have chosen the sign properly. If we choose the orbit (Ic) we get
$(125)-(135)+(735)-(725)-(1246)-(7346)$ where the minus sign of last
two terms is determined by the positive sign of $(125),(735)$. However, 
we find that field $\phi_{46}$ shows up twice with the same sign and
contradicts the toric condition \cite{multiplicity}. 
The same argument shows that the orbit (Id) is not the
correct choice either. This tells us that we should choose the
other combination $6+8+16$.  

For this combination of $6+8+16$, there are two orbits with 16
fields. However, since they are different by only
relabelling $2\leftrightarrow 3$,  
we can choose without loss of generality, for example, the orbit (If).
Starting from this orbit we write down
$$
-\phi_{13} \phi_{35} \phi_{56} \phi_{61}+ \phi_{72} \phi_{25} \phi_{56} \phi_{67}
-\phi_{72} \phi_{24} \phi_{45} \phi_{57}+ \phi_{13} \phi_{34} \phi_{45} \phi_{51}.
$$
Now we need to determine the other orbits. Since $\phi_{13}$ has
appeared twice in orbit (If) already, we must choose orbit (Ic) and (Ia). 
Putting every thing together we get
\bean
W_{I} & =  & [-\phi_{13}\phi_{35} \phi_{56} \phi_{61}
 +\phi_{72}\phi_{25} \phi_{56} \phi_{67}
-\phi_{72}\phi_{24}\phi_{45}\phi_{57}+ \phi_{13}\phi_{34}
\phi_{45}\phi_{51}]\\
& & +[\phi_{35} \phi_{57} \phi_{73}-\phi_{51}\phi_{25}\phi_{12}]
+ [\phi_{12}\phi_{24} \phi_{46} \phi_{61}-\phi_{73} \phi_{34} 
  \phi_{46} \phi_{67}]
\eean
which is exactly the superpotential derived by the unhiggsing in
\eref{superdP4I}.

\EPSFIGURE[ht]{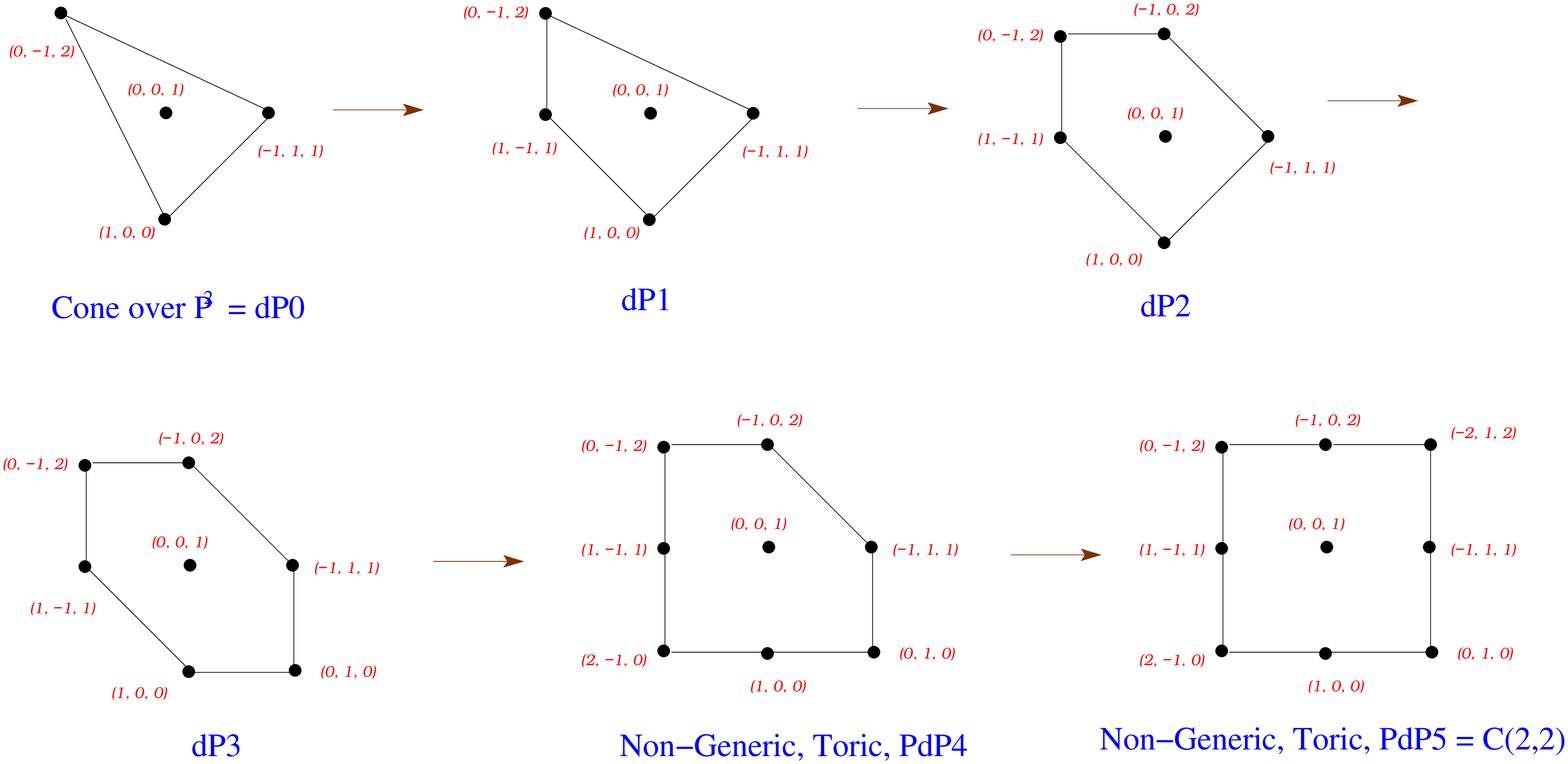,width=17cm}
{The sequence of generic $\IP^1$ blowups from $dP0 = {\cal
O}_{\IP^2}(-3)$ to $dP3$. The last blownups from $dP3$ give
non-generic, toric, $PdP4$ and $PdP5$. 
We have drawn the toric diagrams in a way such that
it is obvious that each blowup corresponds to an addition of a node.
\label{f:blowups}
}
\subsection{Symmetries of $PdP4_{II}$}
Now let us discuss model II with 19 fields. The quiver has the 
symmetry $\IZ_2^{(1)}\times S_3$ where $\IZ_2: 2\leftrightarrow 3$ and
$S_3$ is the symmetric group on the 3 nodes $(1,4,7)$. As before,
we write down the orbits as 
\begin{enumerate}
\item $\{ (125), (135) \}$ + $\{ (425), (435) \}$ + $\{ (725), (735) \}$ 
\item $\{ (1256), (1356) \}$ + $\{ (4256), (4356) \}$ + $\{ (7256), (7356)\}$ 
\item $\{(125436), (135426) \}$ + $\{(125736), (135726) \}$ + $\{(425136), (435126) \}$ + \\ 
 $\{(425736), (435726) \}$ + $\{(725436), (735426) \}$ + $\{(725136), (735126) \}$
\item $\{ (126), (136) \}$ + $\{ (426), (436) \}$ + $\{ (726), (736) \}$
\item $\{(126735), (136725) \}$ + $\{(126435), (136425) \}$ + $\{(426735), (436725) \}$ + \\ 
$\{(426135), (436125) \}$ + $\{(726135), (736125) \}$ + $\{(726435), (736425) \}$, 
\end{enumerate}
where we have divided the action of $\IZ_2^{(1)}$  and $S_3$. Notice
that the number of fields in the orbits are $18,24,72, 18, 72$,  
it is impossible to get the 38 fields needed in the superpotential. 
Again, the quiver symmetry must be broken by the 
superpotential. Let us analyse how the symmetry is broken. 

First we
consider the case that only one symmetry is broken:
(A) $\IZ_2^{(1)}$ is broken and we get orbits with number of fields
$9,12,36$, which can not in any way combine to get 38;
(B) $S_3$ is broken to the cyclic subgroup $\IZ_3$ so that only orbits
(3) and (5) are broken to two parts and we get the numbers
$18,24,36$ which again can not give 38;
(C) $S_3$ is broken to  the subgroup $\IZ_2^{(2)}$ which we can take
to be the action that exchanges nodes $1\leftrightarrow 7$. In this case,
every orbit is broken and we get numbers $12,6,16,8,24$. We have 
five solutions $24+8+6,16+16+6, 16+8+8+6, 12+12+8+6,12+8+6+6+6$ which
give 38.

Now we will try to show that these five solutions can give at most one
consistent result:
\begin{itemize}
\item The $12+8+6+6+6$ case: Only orbits (1) and (3) can be broken to
	provide the number 6. It is easy to see that fields
$\phi_{42},\phi_{43}$ show up three 
times at least, so it is not the correct choice.
\item The $12+12+8+6$ case: number 8 can come only from $\{ (4256),
(4356) \}$ and number 6 can come from $\{ (425), (435) \}$ or $\{
(426), (436) \}$.  
  From the field $\phi_{64}$ we can choose only orbit $\{ (425), (435) \}$
  for the number 8.
  Number 12 can comes from (1)$\{ (125), (135) \}$ +$\{ ((725), (735) \}$;
  (2) $\{ (126), (136) \}$ + $\{ (726), (736) \}$. Since we need two 12,
  every orbit shows up once and only once by considering fields 
  $\phi_{51}, \phi_{61}$. From these arguments, we can write down the
  superpotential uniquely as
  \bean
  W & = & [\phi_{42} \phi_{25} \phi_{56} \phi_{64}-\phi_{43} \phi_{35}
\phi_{56} \phi_{64}] 
  -[ \phi_{42} \widetilde{\phi_{25}} \phi_{54}-\phi_{43}
\widetilde{\phi_{35}} \phi_{54}]\\ 
  & & +[-\phi_{12} \phi_{25} \phi_{51}+ \phi_{13} \phi_{35} \phi_{51}
     +\phi_{72}\widetilde{\phi_{25}}
\phi_{57}-\phi_{72}\widetilde{\phi_{35}} \phi_{57}] \\ 
  & & +[\phi_{12} \phi_{26} \phi_{61}-\phi_{13} \phi_{36} \phi_{61}-
      \phi_{72} \phi_{26} \phi_{67}+\phi_{73} \phi_{36} \phi_{67}]
  \eean
This is a perfect legitimate toric superpotential, but is not the one
found by Seiberg duality from Model I. To see why it is not correct
choice heuristically, notice the term  
  $[\phi_{42} \phi_{25} \phi_{56} \phi_{64}-\phi_{43} \phi_{35}
\phi_{56} \phi_{64}]$ 
  where field $\phi_{56}$  couples to  $\phi_{64}$ two times. In
\cite{multiplicity} we observed 
  that in toric models fields try to couple different field if it is
possible. This may indicate why this is not the right choice.
\item The $16+8+8+6$ case: since number 8 can come only from $\{ (4256),
(4356) \}$, by repeating two times we get that the field $\phi_{64}$
shows up four times, so it is again ruled out.
\item The $16+16+6$ case: number 16 comes only from $\{ (1256),
(1356)\}+\{ ((7256), (7356) \}$. 
Repeating two times will give field $\phi_{56}$ appearing four times,
so it is not allowed either. 
\item The $24+8+6$ case:\\
 number 24 comes from (1) $\{(125436),
(135426) \} 
     +\{(725436), (735426) \}$; (2)$\{(125736), (135726)
\}+\{(725136), (735126) \}$; 
   (3) $\{(425136), (435126) \}+\{(425736), (435726) \}$; (4) 
   $\{(126735), (136725) \}+\{(726135), (736125) \}$; (5) 
   $\{(426735), (436725) \}+\{(426135), (436125) \}$; (6) 
$\{(126435), (136425) \}+ 
  \{(726435), (736425) \}$. As we have showed that $8+6$ can only be
   $\{ (4256), (4356) \}+\{ (425), (435) \}$. However, no matter which
24 we choose, we 
can not satisfy the toric condition: choices (1) and (6) do not give
$\phi_{12}$ appearing two times; choice (2) can not give the
consistent sign for fields $\phi_{12}, \phi_{26}, \phi_{61}$;  
choice (3) has the field $\phi_{64}$ appearing four times and so does
choice (5), for the field $\phi_{54}$; and finally choice (4) can not
give consistent signs for fields $\phi_{12}, \phi_{25}, \phi_{51}$.
\end{itemize}

Having ruled out the case of breaking only one group, 
we consider the case that two symmetry generators are broken: 
\begin{itemize}
\item Only the $\IZ_3$ cyclic symmetry remains: 
	We have orbits with field numbers $9,12, 18$.
	From these three numbers we can not get 38.
\item Only $\IZ_2^{(1)}$ remains: We have orbits with fields $6,8,12$
	which can give 38 by
	$12+12+8+6$, $12+8+6+6+6$, $8+8+8+8+6$ and $8+6+6+6+6+6$. 
It can be shown that there is solution which satisfies the toric
condition, 
for example,
  \bean
  & & (4256)-(4356)-(4\widetilde{25})+(4\widetilde{35})-(125)+(135) \\
  & &+(726)-(736)-(7261\widetilde{35}), (7361\widetilde{25})
  \eean  
  However, for all solutions, we must have at least one of the orbits
with 
8 fields, for example $\{ (4256), (4356) \}$ where field 
  $\phi_{56}$ couples to field $\phi_{64}$ two times. This hints that
is not the   correct choice for this kind of symmetry breaking because
once again fields try to couple to different fields.
\item Only $\IZ_2^{(2)}$ remains: this is similar to the above,
   i.e., from orbits with fields 8 (or two orbits with fields 4), we
will find two fields coupling to each other two times. 
So it hints again that it may not be the correct symmetry.
\item The diagonal $\IZ_2^{(3)}: (2,1) \leftrightarrow (3,7)$: this
will turn out to be
the correct symmetry  preserved by the superpotential.
\end{itemize}

Now we have the correct symmetry $\IZ_2^{(3)}$ to break, let us try to
fix the superpotential. First we write down the orbits as
\begin{itemize}
\item (IIa). (1) $\{ (125), (735) \}$; (2) $\{ (725),(135) \}$; (3)
$\{ (126),(736) \}$; \\
    (4) $\{ (726),(136) \}$; (5) $\{ (425), (435) \}$; (6) $\{ (426),
(436) \}$; 
\item (IIb). (1) $\{ (1256), (7356) \}$; (2) $\{ (7256), (1356) \}$; (3)
$\{ (4256), (4356) \}$; 
\item (IIc). (1) $\{(125436), (735426) \}$; (2) $\{(725436),(135426) \}$;
   (3)  $\{(125736), (735126) \}$; \\ (4) $\{(725136), (135726) \}$; (5)
$\{(425136), (435726) \}$;
   (6) $\{(425736),(435126) \}$;
\item (IId). (1) $\{(126735), (736125) \}$; (2)  $\{(726135),(136725) \}$;
   (3) $\{(126435), (736425) \}$; \\ (4) $\{(726435),(136425) \}$; (5)
$\{(426735),(436125) \}$; (6) $\{(426135), (436725) \}$.
\end{itemize}

There are four ways to get the number 38: $12+12+8+6$, $12+8+6+6+6$,
$8+8+8+8+6$ and $8+6+6+6+6+6$. Let us consider them case by case:
\begin{itemize}
\item The case $8+6+6+6+6+6$: number 8 can come only from (IIb), where
(IIb3) should be excluded 
   because fields $\phi_{56},\phi_{64}$ couple to each other two
times. By relabelling,  
we can fix the number 8 to be the orbit $\{ (7256), (1356) \}$. 
Since $(25,35)$ are doubly degenerate, we need them to show up four
times, so (IIa1), (IIa2) and  (IIa5) must be included.
   Then to complete fields $(42,12)$ we need to include (IIa6) and (IIa3).
   Putting the sign correctly we get the superpotential
   \bean
   W_{II} & = & [(7256)-(1356)]-[(736)-(126)]-[(426)-(436)]
   +[(4\widetilde{25})-(4\widetilde{35})] \\
   & & -[(7\widetilde{25})-(1\widetilde{35})]-[(125)-(735)]
   \eean
   where we use $(25,\widetilde{25})$ to distinguish the fields
$(\phi_{25},\widetilde{\phi_{25}})$. 
This is in perfect agreement with our earlier results (q.v.~\eref{dp4_2}).
\item Other cases: Notice that there
   are three fields $\phi_{72},\phi_{42},\phi_{12}$ which can not
coexist in any orbit. 
To let every field appear twice, we need six terms in the
superpotential. Since all other cases
do not have six terms, the above $8+6+6+6+6+6$ case is the only
allowed choice.
\end{itemize}
\section{Conclusions and Prospects}
The purpose of our writing is to implement the ``unhiggsing
mechanism'' of finding the gauge theories
living on the worldvolume of D-branes probing more general classes of
singularities. In particular, we have addressed blow ups $Y$ of
singularities $X$ (by a $\IP^1$) whose
corresponding probe theories are already known. 

In order to do so, we have developed a
field theoretic
method to obtain the superpotentials once the matter content for the
blowup geometry is at hand.
The approach is based on identifying in each case the unhiggsing
associated to the blown
up 2-cycles\footnote{
  In the case of toric singularities, the $(p,q)$ web techniques
discussed in \cite{webs} and the Inverse Algorithm of \cite{toric}
are very computationally convenient for this purpose.
}
well-known in the literature (cf.~e.g.~\cite{DGM,DM,Uranga}).
Therefore from this standard result that acquisition of VEV's of
spacetime fields is reflected as blow downs in the geometry, while conversely
unhiggsing  corresponds to blowing up, we have devised a
straight-forward algorithm of unhiggsing. The
inputs to the procedure are the matter content and superpotential of
$X$ and the matter content of $Y$; the output is the
superpotential (and hence the full theory) for $Y$.

As applications to our method, we venture into the unchartered waters
of the non-toric higher del Pezzo's. 
Since we know that  each $dPk$ is $dP(k-1)$ blown up at a point and from
the techniques of $(p,q)$-webs, we also know the matter content of the
the higher $dP(k > 3)$ \cite{Hanany:2001py}, it seems  that our unhiggsing
procedure is perfectly adapted to this abovementioned 
problem of finding the full theory for $dP(k > 3)$. Subsequently blow ups
of $dP3$ were constructed along these lines, and the set of all the
toric phases (with equal rank in all their gauge groups) closed under Seiberg
dualities were found. As a confirmation, the inverse procedure of higgsing
in the newly obtained gauge theory (as blow down of the singularity) was then
thoroughly studied and indeed all the toric phases of $dP3$ were retrieved.

The geometry of the unhiggsed theory was
then analyzed in detail. We found there that direct frontal-attack
computation of the moduli space for the theory
gives us a variety which we call {\bf Pseudo $dP4$} or $PdP4$.
This is a toric variety, which is intimately related to $dP4$ in the sense
that it is also $dP3$ blown up at a point, but one which is
non-generic by having non-isolated singularities.
In conclusion, the unhiggsing has provided a new set of theories
supporting the Toric Duality/Seiberg Duality correspondence.

This program was repeated once more for a further blow up. The geometry
of the moduli space is in this case a toric, non-generic
pseudo-$dP5$. It is in fact the generalised conifold $C(2,2)$, which
is a cone over $\IP^2$ blown up at 5 non-generic points. Again Toric
and Seiberg dualities coincide.

Finally we have systematically addressed the symmetries of
these two
new classes of gauge theories along the path of \cite{multiplicity}.
Indeed from
considerations of the global symmetries alone we can obtain the
superpotential by direct
observation and the results are in perfect agreement with the
superpotentials obtained from the unhiggsing method.

We have thus obtained the full theories for some pseudo $dP$'s; 
of immediate concern is of course the question of finding the actual,
generic $dP$'s \cite{appear}. In principle, there are several reasons that 
explain why the direct unhiggsing method does not produce the true
$dP$'s. First, it is possible 
that the quivers found for the higher $dP$'s are incomplete as the
symmetric parts are missing from the $(p,q)$-web method. This
is a general problem when the matter content is calculated just 
from the intersection numbers. Another possibility is that the
four toric phases of $dP3$ are not directly related to the phase
of $dP4$ given by the quiver in \fref{f:dP4I}. In other words,
starting from this phase of $dP4$ we cannot higgs down to the
four phases of $dP3$. In our example, it seems that it is 
 the second reason accounts for our failure. In fact, it can be
seen that the superpotential \footnote{We would
	like to thank Francis Lam
	for collaborating on this part of the calculation.} 
\bean
W & = & [(125)+(735)]+[(1246)+(7346)]+[(1346)]+[(2467)] \\
& & +[(2457)+(2567)]+[(1245)-(5673)]~,
\eean
 where we have grouped terms according to the $Z_2$ symmetry:
$1\leftrightarrow 3,7\leftrightarrow 2,4\leftrightarrow 6$ plus 
charge conjugation, does give the cone over $dP4$ as the moduli space. 
The question now becomes how do we know it is a brane probe
theory if we cannot establish the relationship with  known 
results\footnote{B.~F.~would like
	to thank F.~Cachazo for discussions
about this point.} of $dP3$. 
Work on this issue is in progress \cite{appear}. 

Our
unhiggsing technique thus stands yet another rung on the ladder toward
the solution to general D-brane probe theories upon which we daily
climb. Of course, 
the virtues of the unhiggsing method is appreciable; we are provided
with a technique to address much more general situations than
del Pezzo surfaces, such as arbitrary toric singularities,
or even for singular manifolds with $G_2$ holonomy.

\section*{Acknowledgements}
This Research was supported in part by the patronage of
the CTP and LNS of MIT and the U.~S.~Department of Energy 
under cooperative research agreement \# DE-FC02-94ER40818; the SNS of
IAS under grant PHY-0070928, as well as the Dept.~of Physics at UPenn
under \# DE-FG02-95ER40893.
A.~H.~is also indebted to the Reed Fund Award and a DOE OJI Award.
We are grateful to F.~Lam of M.I.T. for many interesting discussions
and the UROP programme of M.I.T. for providing us with this very
helpful collaborator. B.F. would like to thank Freddy Cachazo for 
valuable conversations. 
YHH would also like to acknowledge M.~Wijnholt for many interesting
discussions and chilling out together; he would in addition like to thank
D.~Grayson and S.~Katz for their invitation to and
wonderful organization of, and especially H.~Schenck and M.~Stillman for their
patient mentoring during the ``Learning Stacks and Computational Methods 
through Problem-Solving Workshop'' at the University of Illinois at
Urbana-Champaign.

\bibliographystyle{JHEP}

\end{document}